\documentclass[11pt]{article}
\usepackage{amsmath,amssymb,amsthm,mathtools}
\usepackage{booktabs}
\usepackage{graphicx}
\usepackage{tikz}
\usetikzlibrary{arrows.meta}
\usepackage[numbers]{natbib}
\usepackage{algorithm}
\usepackage{algpseudocode}
\algrenewcommand{\algorithmicrequire}{Input:}
\algrenewcommand{\algorithmicensure}{Output:}
\usepackage[colorlinks=true,linkcolor=blue,citecolor=blue,urlcolor=blue]{hyperref}
\usepackage{xurl}
\usepackage[nameinlink,capitalize,noabbrev]{cleveref}
\usepackage[margin=1in]{geometry}
\usepackage{authblk}

\makeatletter
\@ifpackagelater{amsmath}{2025/05/31}{%
  \AtBeginDocument{%
    \def\pm@old@displaylabel#1{\cref@old@label@in@display{{#1}}}%
    \ifx\label@in@display@noarg\pm@old@displaylabel
      \def\label@in@display@noarg#1{\cref@old@label@in@display{#1}}%
    \fi
  }%
}{}
\makeatother

\newcommand{\Var}[2][]{\operatorname{Var}_{#1}\left[#2\right]}
\newcommand{\Cov}[2][]{\operatorname{Cov}_{#1}\left[#2\right]}
\newcommand{\indc}[1]{{\mathbf{1}_{\left\{{#1}\right\}}}}
\newcommand{\sfX}{\mathsf{X}}
\newcommand{\sfU}{\mathsf{U}}
\newcommand{\calE}{\mathcal{E}}
\newcommand{\calM}{\mathcal{M}}
\newcommand{\rd}{\mathrm{d}}
\newcommand{\Id}{\mathrm{Id}}
\newcommand{\ep}{\varepsilon}
\newcommand{\R}{\mathbb{R}}
\newcommand{\N}{\mathcal{N}}
\newcommand{\bbP}{\mathbb{P}}
\newcommand{\PP}[2][]{\mathbb{P}_{#1}\left[#2\right]}

\newcommand{\Indc}[1]{{\mathbf{1}\left\{{#1}\right\}}}
\newcommand{\Gap}{\operatorname{Gap}}
\newcommand{\osc}{\operatorname{osc}}
\newcommand{\oracle}{{\mathrm{or}}}
\newcommand{\PM}{{\mathrm{PM}}}
\newcommand{\argmin}{\operatorname{argmin}}
\newcommand{\tran}{^\intercal}
\newcommand{\Rom}[1]{\uppercase\expandafter{\romannumeral #1\relax}}
\newcommand{\widebar}[1]{{%
  \mkern1.5mu
  \overline{\mkern-1.5mu#1\mkern-1.5mu}%
  \mkern1.5mu
}}

\newcommand{\numberthis}{\addtocounter{equation}{1}\tag{\theequation}}
\newtheorem{theorem}{Theorem}[section]
\newtheorem{lemma}[theorem]{Lemma}

\newtheorem{remark}[theorem]{Remark}

\begin{document}

\title{Pseudo-marginal Sampling for Selective Inference}
\author[1]{Jiali He}
\affil[1]{Department of Statistics and Finance, University of Science and Technology of China}
\author[2]{Sifan Liu}
\affil[2]{Department of Statistical Science, Duke University}
\date{\today}

\maketitle
\begin{abstract}
Inference after adaptive selection must account for selection bias. One approach is to condition on the selected event, but for complex selection procedures, the resulting conditional distribution can be analytically intractable and require sampling-based methods. Rejection sampling provides a natural baseline: it repeatedly draws from a proposal and retains only samples that reproduce the observed selection. This can be prohibitively expensive when the selection event is rare. We address this computational bottleneck with a pseudo-marginal Markov chain Monte Carlo algorithm that samples from the selective distribution. Using a weak Poincar\'e inequality, we analyze its computational efficiency relative to rejection sampling and identify when it can offer substantial gains. In the classical file-drawer problem, we derive precise computational costs for both methods, illustrating exponential improvement as the selection event becomes increasingly rare. In numerical experiments involving complex selection procedures without tractable conditional distributions, the pseudo-marginal sampler is often hundreds of times more efficient than rejection sampling, while the resulting tests achieve nominal selective Type I error.
\end{abstract}

\section{Introduction}

Modern statistical analyses use data to decide which questions to ask as well as how to answer them. A regression analysis may choose variables after tuning its regularization parameter by cross-validation~\citep{markovic2017unifying}; a changepoint analysis may select both the number and locations of changes~\citep{hyun2021changepoint}. These adaptive choices affect the distributions of the statistics subsequently used for inference and can invalidate conventional tests and confidence intervals, typically leading to overconfident claims~\citep{berk2013valid}. Selective inference addresses this problem by conditioning on the selected outcome~\citep{fithian2014optimal,lee2016exact}. Calibrating tests and confidence intervals then requires characterizing or sampling from the resulting conditional distribution.

For some selection procedures, the conditional distribution admits a tractable analytical characterization. Examples include polyhedral selection events arising from lasso variable selection and related procedures~\citep{lee2016exact,tibshirani2016exact,taylor2018post,hyun2018exact}. Under an exponential family model and with suitable additional conditioning, the conditional distribution is univariate and truncated to an interval, allowing for exact inference. In practice, however, increasingly sophisticated analysis pipelines often involve adaptive hyperparameter tuning and multiple stages of selection, making the selection event difficult to characterize analytically.
Selection procedures may also involve randomization, whether introduced to improve statistical power~\citep{tian2018selective} or arising from random subsampling during cross-validation. Marginalizing over this randomization when conditioning on the selected outcome can further complicate the selective distribution.

When the selective distribution is analytically intractable, one can treat the selection procedure as a simulator and run it on simulated data to determine whether the observed selection is reproduced. These simulations can be used to estimate the selection probability~\citep{liu2022black} or to learn a transport map from the conditional distribution to the pre-selection distribution without explicitly characterizing the selection event~\citep{liu2025flexible}. When fresh randomization is used on each run, repeated simulations also implicitly average over its distribution. However, these simulation-based approaches critically depend on the ability to generate data that satisfy the selection event.

Rejection sampling provides a direct way to sample from the selective distribution: draw independently from the pre-selection distribution, rerun the selection procedure, and retain a simulated dataset if it reproduces the observed selection~\citep{liu2025flexible,freidling2026selective}. The accepted datasets are independent and identically distributed (i.i.d.) draws from the selective distribution and can be used for inference; for example, the observed test statistic can be compared with quantiles of the corresponding empirical distribution. Each proposal, however, requires rerunning the entire selection procedure. When the selection event is rare, most proposals are rejected, making rejection sampling prohibitively expensive.

To address this bottleneck, we propose a Markov chain Monte Carlo (MCMC) approach. At each step, the sampler proposes a local move from its current state and reruns the selection procedure to check whether the observed selection is reproduced. A suitable Metropolis--Hastings (MH) acceptance rule yields a Markov chain whose stationary distribution is the desired selective distribution. Unlike rejection sampling, which draws anew from the pre-selection distribution on every attempt, the Markov chain can continue exploring regions favored by selection. Local proposals may therefore be accepted more often and, when selection is rare, improve computational efficiency relative to rejection sampling.

This algorithm is an instance of pseudo-marginal (PM) MCMC~\citep{andrieu2009pseudo}, which uses nonnegative unbiased estimates of intractable density factors in a Metropolis--Hastings algorithm. Here, the intractable factor is the probability of reproducing the observed selection conditional on the proposed state, and the unbiased estimate is the indicator of whether the observed selection is reproduced. PM can be viewed as an auxiliary-variable MCMC method in which the augmented state includes the randomization variables used in the simulator. Targeting this augmented distribution implicitly marginalizes over the randomization, yielding the target selective distribution as the stationary marginal. Related simulation-based MCMC methods are used in approximate Bayesian computation~\citep{marjoram2003markov,marin2012approximate,fearnhead2012constructing,lee2014variance}, in sampling from doubly intractable distributions~\citep{murray2006mcmc,yuan2026markov}, in Bayesian inference for state-space models via particle MCMC~\citep{andrieu2010particle}, and in Bayesian inference from privatized data~\citep{ju2022data}.

The pseudo-marginal framework provides a general construction, but it does not by itself guarantee computational efficiency. Indeed, naive rejection sampling can also be viewed as a pseudo-marginal algorithm with independent proposals. Local proposals may increase the acceptance probability, but greater acceptance need not improve sampling efficiency because the resulting chain is autocorrelated. The central question is therefore how to choose a transition kernel and under what conditions it outperforms rejection sampling at the same computational cost, especially when the selection event is rare.

Our analysis clarifies when local proposals can overcome the rarity of selection. Building on the weak Poincar\'e comparison framework of \citet{andrieu2022comparison}, we derive conditions under which PM-MCMC achieves a given accuracy at lower computational cost than rejection sampling. Intuitively, if selection is much more likely in a typical region of the selective distribution than under pre-selection sampling, local proposals can sample more efficiently within that region. As we will see, the selective inference examples studied here exhibit this favorable pattern. Conversely, if selection remains unlikely even in a typical region of the selective distribution, local exploration offers less benefit. Thus, the computational efficiency of the PM sampler depends not simply on the overall rarity of selection but also on how the selection probability varies across the state space.

Our main contributions are as follows.
\begin{itemize}
    \item We propose a pseudo-marginal MCMC method for sampling from selective distributions. It handles complex randomized selection procedures that can be rerun without evaluating their selection probabilities, and allows the conditioning event to be tailored to the inferential target by specifying which selection outputs must match the observed ones.
    \item We compare the computational efficiency of the PM sampler with that of rejection sampling, first through a general comparison using the weak Poincar\'e framework~\citep{andrieu2026weak}.
    We then consider the file-drawer model, a classical model of publication bias~\citep{rosenthal1979file,iyengar1988selection}, and derive an exact efficiency comparison among PM, correlated PM~\citep{deligiannidis2018correlated}, and rejection sampling. As the selection event becomes increasingly rare, PM and correlated PM achieve exponential improvements over rejection sampling.
    \item We evaluate the proposed method numerically. In terms of sampling efficiency, it achieves several-hundred-fold gains relative to rejection sampling at the same computational cost.
    It also correctly controls the selective Type I error and achieves higher empirical power than data splitting. Our experiments also show that ignoring hyperparameter tuning can inflate Type I error.
\end{itemize}

The remainder of the paper is organized as follows.
Section~\ref{sec:selective-inference} introduces the selective inference problem and the sampling task.
Section~\ref{sec:setup} develops the PM sampler and compares it with rejection sampling and an oracle sampler that has access to the exact selection probability.
Section~\ref{sec:general_theory} analyzes sampling efficiency using weak Poincar\'e inequalities. For the file-drawer problem, we derive sample-complexity rates for PM and correlated PM as selection becomes increasingly rare.
Section~\ref{sec:experiments} presents numerical experiments on sampling efficiency and selective inference tasks for which the selection event is analytically intractable.
Section~\ref{sec:discussion} concludes with a discussion of future directions.
Proofs and additional experimental details appear in the appendices.

\section{Selective inference and the sampling problem}
\label{sec:selective-inference}

\subsection{Inference conditional on selection}
\label{sec:selinf}

Data-dependent decisions, such as selecting a model or choosing whether to report a result, affect the subsequent inference. Let $Y\in\R^n$ denote the data and $W\in\R^m$ any external randomization used by the selection procedure $\widehat M$. Given observed realizations $Y=y$ and $W=w$, the selected output is $M=\widehat M(y,w)\in\calM$, where $\calM$ is the set of possible outputs. This output determines the target of inference: for example, if $\widehat M$ selects variables, the target may be their coefficients. We write $\theta_M$ for the parameter associated with this target. The framework also covers selective reporting or detection, where a result is reported only when specified criteria are met; the distribution of reported data is then conditional on those criteria.

Because selection changes the distribution of the data being analyzed, inference procedures that ignore this selection effect may be invalid. Selective inference addresses this by conditioning on the selected output~\citep{fithian2014optimal,lee2016exact}. For example, when testing $H_0:\theta_M=\theta_0$, suppose the selective Type I error
\begin{align}\label{eq:selective_type1_error}
\PP[\theta_0]{\text{reject } H_0 \mid \widehat M = M}
\end{align}
is at most $\alpha\in(0,1)$ for every possible selected output $M\in \calM$. Averaging over the possible outputs then gives a marginal Type I error of at most $\alpha$.

Let $T=T(Y)\in\R^d$ denote the statistic used to test $H_0:\theta_M=\theta_0$. We first assume that selection depends on the data and randomization only through $(T,W)$. More generally, if $\widehat M$ also depends on a statistic $S=S(Y)$, we condition on its observed value $S=s$. With $s$ held fixed, selection depends only on $(T,W)$, and we write the selection procedure as $\widehat M(T,W)$. We also assume that the conditional model has no nuisance parameters under $H_0$. In an exponential family, this may be achieved by including sufficient statistics for the nuisance parameters in $S$ and conditioning on them. Whenever we condition on $S=s$, the distribution of $(T,W)$ is understood as conditional on $S=s$. See Section~\ref{sec:experiments} for examples of the conditioning statistics used in our applications.

Let $p_{\theta_0}(t,w)$ be the joint density of $(T,W)\in\R^d\times\R^m$ under $H_0:\theta_M=\theta_0$, before conditioning on selection. As above, this density is conditional on $S=s$ when $S$ is used. For an output $M$ with positive selection probability, the density of $T$ conditional on selecting $M$ is
\begin{align}\label{eq:selective_density_t}
p_{\theta_0}(t\mid\widehat M=M)
&=
\frac{\displaystyle\int_{\R^m}
    \indc{\widehat M(t,w)=M}\,p_{\theta_0}(t,w)\,\rd w}
{\displaystyle\int_{\R^d}\int_{\R^m}
    \indc{\widehat M(u,v)=M}\,p_{\theta_0}(u,v)\,\rd v\,\rd u}
=
\frac{p_{\theta_0}(t)\,L_{\theta_0}(t)}
{\bbP_{\theta_0}[\widehat M(T,W)=M]},
\end{align}
where
\[
L_{\theta_0}(t)
:=
\PP[\theta_0]{\widehat M(t,W)=M\mid T=t}
\]
is the probability of selecting $M$ conditional on $T=t$, and $p_{\theta_0}(t)$ is the marginal density of $T$ under the null before selection. Thus selection reweights the pre-selection density $p_{\theta_0}(t)$ by $L_{\theta_0}(t)$; the normalizing constant is the overall selection probability $\PP[\theta_0]{\widehat M(T,W)=M}$. When the selection rule is fixed and the conditional distribution of $W$ given $T=t$ is parameter-free, $L_{\theta_0}(t)$ does not depend on $\theta_0$. We retain the subscript to cover the general case.

The conditional law in~\eqref{eq:selective_density_t} provides the basis for selective inference. If it can be evaluated analytically, a selective $p$-value can be computed directly; for a univariate $T$ with observed value $t=T(y)$, a one-sided $p$-value is $\PP[\theta_0]{T\geq t\mid\widehat M=M}$. Polyhedral selection events make such calculations possible for lasso selection, sequential regression, and generalized-lasso changepoint detection after suitable conditioning~\citep{lee2016exact,tibshirani2016exact,hyun2018exact}. When the selection event or conditional law is intractable, this analytical route is unavailable.

\paragraph{Running example: file-drawer problem.}
\label{sec:file_drawer}
The classic file-drawer problem provides a simple example in which the effect of selection can be seen explicitly~\citep{rosenthal1979file,iyengar1988selection}. Let $T\sim\N(\theta_0,1)$ and $W\sim\N(0,1)$ be independent. A result is reported only when a weighted combination of $T$ and $W$ exceeds a threshold:
\begin{align*}
\big\{\widehat M(T,W)=M\big\} = \big\{\sqrt{a}\,T+\sqrt{1-a}\,W>b\big\}.
\end{align*}
Here $a\in(0,1)$ controls the randomization level, and $b\in\R$ is the reporting threshold. Write $\varphi$ for the standard normal density and $\widebar\Phi(u)=1-\Phi(u)$ for the corresponding survival function. The selection probability at each value $T=t$ is
\[
    L_{\theta_0}(t)
    =\PP[\theta_0]{\sqrt{a}\,t+\sqrt{1-a}\,W>b\mid T=t}
    =\widebar\Phi\Big(\frac{b-\sqrt a t}{\sqrt{1-a}}\Big),
\]
which does not depend on $\theta_0$. Integrating over $T$ gives the overall probability of selecting $M$:
\begin{align*}
    \PP[\theta_0]{\widehat M(T,W)=M}=\widebar\Phi(b-\sqrt a\theta_0).
\end{align*}
Substituting these probabilities into~\eqref{eq:selective_density_t} gives the density of $T$ conditional on selection:
\[
    p_{\theta_0}(t\mid \widehat M=M)
    =
    \frac{\varphi(t-\theta_0)\widebar\Phi(\frac{b-\sqrt a t}{\sqrt{1-a}}) }
         {\widebar\Phi(b-\sqrt a\theta_0)}.
\]
As $b$ increases, the selection probability $\widebar\Phi(b-\sqrt a\theta_0)$ tends to zero. In Section~\ref{sec:file-drawer-analysis}, we use this model to compare the sampling efficiency of rejection sampling and pseudo-marginal methods in this rare-selection regime.

\subsection{Sampling from the selective distribution}
\label{sec:sampling-selective}

When the selective distribution~\eqref{eq:selective_density_t} cannot be evaluated analytically, we can use Monte Carlo samples for inference. To see how, let $f:\R^d\to\R$ be a test function for which larger values of $f(T)$ provide stronger evidence against $H_0$. For the observed statistic $t$, define the selective $p$-value as
\begin{align}\label{eq:selective_p_value}
    \phi(t) = \PP[\theta_0]{f(T) \geq f(t) \mid \widehat M=M }.
\end{align}
For every $\alpha\in(0,1)$, this $p$-value obeys
\begin{align*}
    \PP[\theta_0]{\phi(T) \leq \alpha \mid \widehat M=M} \leq \alpha.
\end{align*}
Thus, rejecting $H_0$ when $\phi(T)\leq\alpha$ controls the selective Type I error~\eqref{eq:selective_type1_error} at level $\alpha$. Given draws \(T^{(i)} \sim p_{\theta_0}(\cdot \mid \widehat M = M)\), $i=1,\ldots,N$, the selective $p$-value~\eqref{eq:selective_p_value} can be estimated by
\begin{equation}
    \widehat{\phi}(t)
    =
    \frac{1}{N}
    \sum_{i=1}^{N}
    \Indc{f(T^{(i)}) \geq f(t)}.
    \label{eq:selective-pvalue-mc}
\end{equation}

The challenge is then to obtain samples from the selective distribution~\eqref{eq:selective_density_t}. For tractable Gaussian polyhedral events, specialized samplers such as hit-and-run can be used~\citep{fithian2014optimal}. When the selection event is intractable, rejection sampling is available if we can simulate from the pre-selection law and evaluate $\widehat M(t,w)$ at any $(t,w)$. It generates proposals $(T,W)\sim p_{\theta_0}(t,w)$ and retains $T$ when $\widehat M(T,W)=M$. Each proposal requires rerunning the selection procedure to check this event. Its acceptance probability is the overall selection probability under the null, $\PP[\theta_0]{\widehat M(T,W)=M}$. The expected number of selection-procedure evaluations per accepted draw is therefore the reciprocal of this probability and can be large when selection is rare.

This inefficiency arises because rejection sampling starts afresh from the pre-selection distribution on every attempt, even after finding a draw that satisfies selection. This motivates the MCMC approach developed next, which proposes local moves from its current state.

\section{Pseudo-marginal sampling from selective distributions}
\label{sec:setup}

\subsection{The pseudo-marginal sampler}
We now construct a pseudo-marginal MCMC sampler for the selective distribution~\eqref{eq:selective_density_t}. Writing $x$ for a generic value of the statistic and suppressing the dependence on $\theta_0$ and $M$, we express the target density as
\begin{align*}
    \pi(x)=\frac{p(x)L(x)}{C},
\end{align*}
where $x$ belongs to a Borel subset $\sfX\subseteq\R^d$ equipped with its Borel $\sigma$-algebra $\mathcal B(\sfX)$, and $C=\int_{\sfX}p(x)L(x)\,\rd x>0$. In the selective-inference setting, $p$ is the pre-selection density, $L(x)$ is the conditional selection probability, and $C$ is the overall selection probability. The sampler also applies when $p$ and $L$ have other interpretations. We assume:
\begin{enumerate}
    \item $p$ is a probability density on $\sfX$ that can be evaluated up to a normalizing constant.
    \item $L$ has the representation
    \begin{align}\label{eq:def_L}
        L(x)=\PP[U\sim m]{(x,U)\in A},
    \end{align}
    where $m$ is a probability measure on $(\sfU,\mathcal B(\sfU))$, with $\sfU\subseteq\R^m$ a Borel set, and $A\in\mathcal B(\sfX)\otimes\mathcal B(\sfU)$. We can simulate $U\sim m$ and query $\Indc{(x,u)\in A}$ for any $(x,u)\in\sfX\times\sfU$.
\end{enumerate}

The construction requires neither evaluating $L(x)$ nor knowing $C$. In selective inference, $U$ represents the randomness used to generate the randomization variable $W$; we assume that $U$ is independent of $Y$ before conditioning. For instance, when $W$ is independent of $Y$, we can take $U=W$ and $A=\{(t,w):\widehat M(t,w)=M\}$. When $W$ depends on $Y$, the same formulation applies if its conditional law given $T=t$ can be simulated as $F(t,U)$ for a measurable function $F$, with $U\sim m$ independent of $Y$. We then take $A=\{(t,u):\widehat M(t,F(t,u))=M\}$.

The event representation~\eqref{eq:def_L} yields a nonnegative unbiased estimator of $L(x)$: for $U\sim m$, the indicator $\Indc{(x,U)\in A}$ has expectation $L(x)$. In selective inference, evaluating this indicator amounts to running the selection procedure and checking whether its output is $M$. This places the target within the pseudo-marginal (PM) MCMC framework~\citep{beaumont2003estimation,marjoram2003markov,andrieu2009pseudo}. We may refer to $p$ and $L$ as the ``prior'' and ``likelihood,'' respectively, as PM-MCMC is typically applied in Bayesian settings, although we do not impose a Bayesian interpretation here.

To construct the sampler, we augment $x$ with the auxiliary variable $u$ and define a joint target measure on $\sfX\times\sfU$ by
\begin{equation*}
    \widetilde\pi(\rd x,\rd u)
    =
    C^{-1}\, p(x)\,\rd x\, m(\rd u)\, \Indc{(x,u)\in A}.
\end{equation*}
This is the joint law of independent $X\sim p$ and $U\sim m$, conditioned on $(X,U)\in A$. By~\eqref{eq:def_L}, integrating over $u$ gives the $x$-marginal $C^{-1}p(x)L(x)\,\rd x=\pi(x)\,\rd x$. We can therefore sample from the desired distribution $\pi$ by constructing a Markov chain with invariant measure $\widetilde\pi$ and retaining its $x$-coordinate.

Let $Q$ be a Markov transition kernel on $(\sfX,\mathcal{B}(\sfX))$ that is reversible with respect to the probability measure with density $p$:
\begin{align*}
    p(x)\,\rd x\, Q(x,\rd x') = p(x')\,\rd x'\, Q(x',\rd x).
\end{align*}
From a current state $x$ in the support of $\pi$, we draw independently
\begin{align*}
    x'\sim Q(x,\cdot),\qquad u'\sim m(\cdot).
\end{align*}
We accept $x'$ if $(x',u')\in A$; otherwise, we retain $x$ as the next state. Algorithm~\ref{alg:pm} summarizes this transition. The auxiliary variable $u$ need not be stored, since the acceptance decision depends only on the proposed pair $(x',u')$. In selective inference applications, the observed statistic $t$ provides a natural initialization for the chain: $L(t)>0$ and hence $\pi(t)>0$, provided that the support of $p_\theta$ does not exclude $t$. The algorithm allows local moves in $x$ while refreshing $u$ independently; Section~\ref{sec:product-event-comparison} explains when this choice can improve efficiency over rejection sampling.

\begin{algorithm}[htbp]
\caption{Pseudo-marginal transition}
\label{alg:pm}
\begin{algorithmic}[1]
    \Require Current state $x$ in the support of $\pi$; query access to the indicator of $A$; a Markov kernel $Q$ reversible with respect to $p$; a sampler for $m$.
    \Ensure Next state $x_{\mathrm{next}}$.
    \State Draw independently $x'\sim Q(x,\cdot)$ and $u'\sim m$.
    \State Query the event indicator $\delta=\Indc{(x',u')\in A}$.
    \State \Return $\delta x' + (1-\delta)x$.
\end{algorithmic}
\end{algorithm}

Averaging the acceptance indicator over the proposed $u'\sim m$ gives the transition kernel for $x$:
\begin{align}\label{eq:marginal-kernel}
    K_{\PM}(x,\rd x') &= Q(x, \rd x') L(x') + \Big(1 - \int_{\sfX} Q(x, \rd y) L(y) \Big) \delta_{x}(\rd x').
\end{align}
The first term accounts for accepted proposals, and the second for rejected proposals. The following theorem establishes reversibility with respect to $\pi$.

\begin{theorem}[Reversibility]\label{thm:reversibility}
    If $L$ satisfies~\eqref{eq:def_L} and $Q$ is reversible with respect to $p$, then the transition kernel $K_{\PM}$ in~\eqref{eq:marginal-kernel} is reversible with respect to $\pi$.
\end{theorem}
A proof is given in Appendix~\ref{prf:reversibility}.
We have a few remarks regarding Algorithm~\ref{alg:pm} and its relation to the usual pseudo-marginal and correlated pseudo-marginal algorithms.

\begin{remark}[Equivalence to the usual PM algorithm]\label{rmk:joint_pm_equivalence}
    Algorithm~\ref{alg:pm} is equivalent, in its $x$-marginal transition, to the usual PM algorithm~\citep{andrieu2009pseudo} with a single indicator estimator. To see this, let $q$ be a proposal density for $x$ and write $\widehat L(x,u)=\Indc{(x,u)\in A}$. From a feasible current state $(x,u)\in A$, the usual PM algorithm proposes independently $x'\sim q(x,\cdot)$ and $u'\sim m$ and accepts $(x',u')$ with probability
    \[
    1 \wedge \frac{p(x')q(x',x) \widehat L(x', u') }{p(x) q(x, x')  \widehat L(x,u)}.
    \]
    If $Q$ is the Metropolis--Hastings kernel with proposal $q$ and target density $p$, this acceptance rule gives the same transition kernel for $x$ as Algorithm~\ref{alg:pm}; see Appendix~\ref{prf:joint_pm_equivalence} for the derivation. The equivalence relies on the binary-valued estimator $\widehat L(x,u)$. We use the formulation in Algorithm~\ref{alg:pm} because it simplifies the analysis below.
\end{remark}

\begin{remark}[Use of multiple auxiliary variables]\label{rmk:multiple_auxiliary}
    More generally, the usual PM algorithm can use the estimator $\widehat L(x, u_{1:B})=\frac1B\sum_{b=1}^B \Indc{(x,u_b)\in A}$, where $u_1,\ldots,u_B\stackrel{\mathrm{i.i.d.}}{\sim} m$ and $B\geq 1$. We focus on $B=1$ for two reasons. First, the equivalence in the preceding remark gives the simpler formulation in Algorithm~\ref{alg:pm}, which facilitates our analysis. Second, in the context of approximate Bayesian computation (ABC), \citet{bornn2017use} show that $B=1$ is nearly optimal when asymptotic variance is adjusted for simulation cost. Increasing $B$ also does not necessarily resolve convergence difficulties: under the local-proposal and tail conditions of \citet{lee2014variance}, the corresponding PM kernels are not geometrically ergodic for any finite $B$.
\end{remark}

\begin{remark}[Correlated pseudo-marginal]\label{rmk:correlated_pm}
    Algorithm~\ref{alg:pm} refreshes the auxiliary variable independently at each iteration. Alternatively, we can retain $u$ as part of the state and propose independently $x'\sim Q(x,\cdot)$ and $u'\sim R(u,\cdot)$, where $R$ is reversible with respect to $m$. This allows dependence between successive auxiliary variables. Since $Q$ is reversible with respect to the measure with density $p$, the joint proposal is reversible with respect to $p(x)\,\rd x\,m(\rd u)$. We therefore accept $(x',u')$ if it belongs to $A$ and otherwise retain $(x,u)$, obtaining a chain reversible with respect to $\widetilde\pi$. This gives the correlated pseudo-marginal (CPM) algorithm~\citep{deligiannidis2018correlated} in our setting. Our main theoretical analysis focuses on independent auxiliary refreshment; we study CPM in the file-drawer example in Section~\ref{sec:file-drawer-analysis} and evaluate it numerically in Section~\ref{sec:experiments}.
\end{remark}

\subsection{Comparison with the rejection and oracle samplers}
\label{sec:product-event-comparison}

Two samplers are particularly relevant for comparison: the rejection sampler and the oracle sampler.
The rejection sampler can be viewed as the PM algorithm with the independent proposal $Q(x, \rd x')=p(\rd x')$: at each iteration, it independently draws $x'\sim p$ and $u'\sim m$ and accepts $x'$ if $(x',u')\in A$. The difference is that PM-MCMC retains the current state upon rejection, whereas rejection sampling discards the failed proposal and tries again.
Substituting $Q(x,\rd x')=p(\rd x')$ into~\eqref{eq:marginal-kernel} gives the transition kernel
\begin{equation*}
    K_{\rm R}(x,\rd x')
    =
    C\pi(\rd x')+(1-C)\delta_x(\rd x'),
\end{equation*}
so the acceptance probability is $C$.

The oracle kernel, also called the marginal kernel, assumes access to the exact event probability $L(x)$. It draws a proposal $x'\sim Q(x,\cdot)$ from the same $p$-reversible kernel and accepts it with probability $1\wedge\frac{L(x')}{L(x)}$. Its transition kernel is
\begin{equation}
        K_{\rm or}(x,\rd x')
        =
        Q(x,\rd x')
        \Big(
            1\wedge\frac{L(x')}{L(x)}
        \Big)+
        \Big[
            1-\int Q(x,\rd y)
            \Big(
                1\wedge\frac{L(y)}{L(x)}
            \Big)
        \Big]\delta_x(\rd x').
    \label{eq:oracle-kernel}
\end{equation}
Comparing $K_{\rm PM}$ in~\eqref{eq:marginal-kernel} with $K_{\rm or}$ in~\eqref{eq:oracle-kernel} shows that their stationary flows satisfy, for $x\neq x'$,
\begin{equation}
    \pi(\rd x)K_{\rm PM}(x,\rd x')
    =
    \max \big\{L(x),L(x')\big\}\,
    \pi(\rd x)K_{\rm or}(x,\rd x').
    \label{eq:exact-flow-comparison}
\end{equation}

We compare these kernels through their spectral gaps, which quantify convergence and control Monte Carlo variance. For a $\pi$-reversible kernel $K$ whose Markov operator is positive semidefinite on $L^2(\pi)$, define its spectral gap by
\begin{equation*}
\Gap(K)=\inf_{\Var[\pi]{f}>0} \frac{\calE_K(f,f)}{\Var[\pi]{f}},
\quad
\calE_K(f,f)=\frac12\int \big(f(y)-f(x)\big)^2\pi(\rd x)K(x,\rd y),
\end{equation*}
where $\calE_K$ is the Dirichlet form and the infimum is over $f\in L^2(\pi)$ with positive variance. Under these assumptions, a positive gap guarantees geometric convergence in $\chi^2$-divergence whenever the initial law has finite $\chi^2$-divergence relative to $\pi$~\citep[Theorem~2.1]{roberts1997geometric}. It also bounds the asymptotic variance of stationary sample averages uniformly over functions with unit variance under $\pi$; a larger gap gives a tighter bound~\citep{roberts2008variance}.

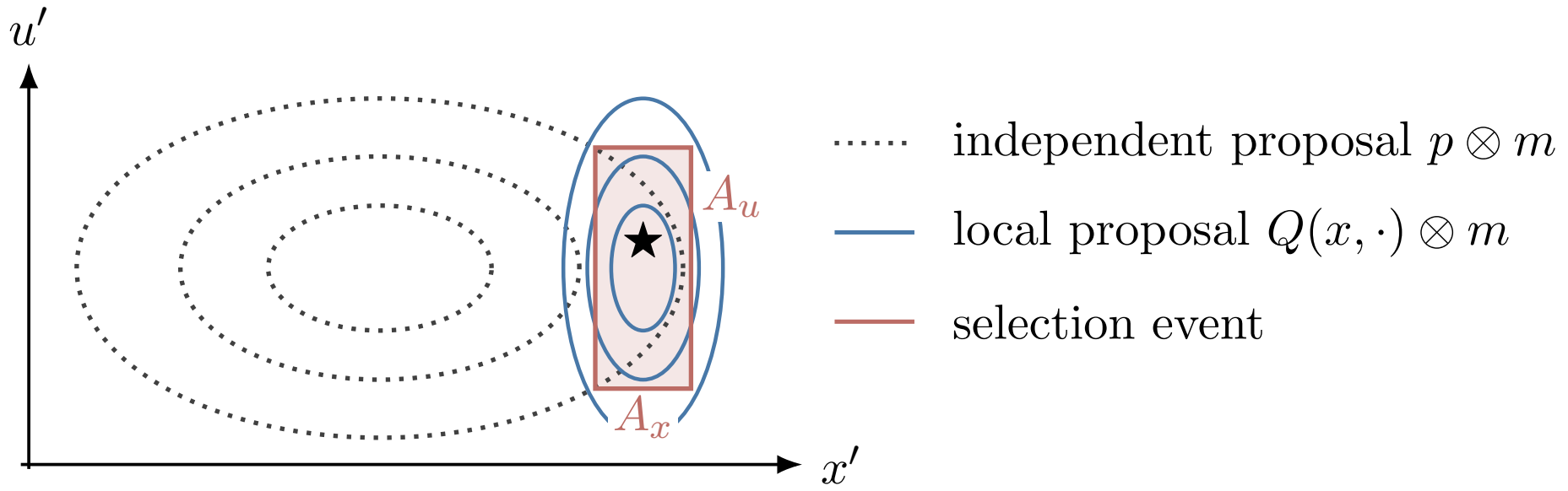
\begin{figure}[t]
    \centering
    \begin{tikzpicture}[x=1.25cm,y=0.70cm,font=\normalsize]
        \definecolor{localblue}{HTML}{4878A8}
        \definecolor{eventorange}{HTML}{BC6C65}
        \fill[eventorange!16] (1.35,-1.35) rectangle (1.95,1.35);
        \foreach \r in {0.70,1.25,1.90} {
            \draw[black!75,dotted,line width=1pt] (0,0) circle[radius=\r];
        }
        \foreach \rx/\ry in {0.2/0.70,0.35/1.25,0.5/1.90} {
            \draw[localblue,line width=0.85pt] (1.65,0) ellipse[x radius=\rx,y radius=\ry];
        }
        \draw[eventorange,line width=1pt] (1.35,-1.35) rectangle (1.95,1.35);
        \node[black,inner sep=0pt] at (1.65,0.30) {\small$\bigstar$};

        \draw[-{Latex[length=2mm]},line width=0.7pt] (-2.25,-2.20) -- (2.65,-2.20) node[right] {$x'$};
        \draw[-{Latex[length=2mm]},line width=0.7pt] (-2.20,-2.25) -- (-2.20,2.30) node[above] {$u'$};

        \node[eventorange,below,fill=white,inner sep=1pt] at (1.65,-1.38) {$A_x$};
        \node[eventorange,right,fill=white,inner sep=1pt] at (1.98,0.80) {$A_u$};
        \draw[black!75,dotted,line width=1pt] (2.85,1.40) -- (3.35,1.40);
        \node[anchor=west] at (3.48,1.40) {independent proposal $p\otimes m$};
        \draw[localblue,line width=0.95pt] (2.85,0.40) -- (3.35,0.40);
        \node[anchor=west] at (3.48,0.40) {local proposal $Q(x,\cdot)\otimes m$};
        \draw[eventorange,line width=0.95pt] (2.85,-0.60) -- (3.35,-0.60);
        \node[anchor=west] at (3.48,-0.60) {selection event};
    \end{tikzpicture}
    \caption{Independent and local proposals for the product event $A=A_x\times A_u$, shown by the red rectangle. The star ($\star$) marks the current state $(x,u)\in A$. Dotted gray contours represent the independent proposal $p\otimes m$ used in rejection sampling; solid blue contours represent the local PM proposal $Q(x,\cdot)\otimes m$. Both proposals draw $u'$ independently from $m$, but the local proposal can place more mass in $A_x$ by proposing $x'$ near the current $x$. When $A_x$ is rare under $p$, this can increase the probability of proposing a pair in $A$.}
    \label{fig:product-rarity}
\end{figure}

\paragraph{Illustration on product events}
To build intuition, we consider the situation where $A=A_x\times A_u$ is a product event, in which case $X$ and $U$ are independent under the joint target $\widetilde\pi$. Denote $\rho_x=p(A_x)>0$ and $\rho_u=m(A_u)>0$. Then $L(x)=\rho_u\indc{x\in A_x}$, $C=\rho_x\rho_u$, and $\pi=p(\cdot\mid A_x)$.
See a schematic illustration in Figure~\ref{fig:product-rarity}.

In this case, the oracle kernel is $K_{\rm or}=Q_{A_x}$, obtained by restricting $Q$ to $A_x$. The PM kernel additionally rejects a proposal when the independent auxiliary draw falls outside $A_u$, an event with probability $1-\rho_u$. Hence $K_{\rm PM}=(1-\rho_u)\Id+\rho_u Q_{A_x}$, which implies
\[
\Gap(K_{\rm PM})=\rho_u\, \Gap(Q_{A_x}).
\]
Thus, the PM kernel inherits the oracle kernel's spectral gap, reduced by a factor of $\rho_u$ because of auxiliary rejection. Rejection sampling has a gap of $\Gap(K_{\rm R})=C=\rho_x\rho_u$; its gap incurs a factor of $\rho_x$ from the rarity of $A_x$ and a factor of $\rho_u$ from the rarity of $A_u$.

The ratio of the spectral gaps of PM and rejection sampling is
\[
\frac{\Gap(K_{\rm PM})}{\Gap(K_{\rm R})}
=\frac{\Gap(Q_{A_x})}{\rho_x}.
\]
Hence, PM has a larger spectral gap precisely when $\Gap(Q_{A_x})>\rho_x$. Here, $\rho_x$ is the spectral gap of the sampler that independently proposes $x\sim p$ and accepts only if $x\in A_x$. Thus, in terms of spectral gap, PM is more efficient than rejection sampling when $Q_{A_x}$ explores $A_x$ more efficiently than independent proposals from $p$, as illustrated in Figure~\ref{fig:product-rarity}.

At the other extreme, suppose $A_x=\sfX$, so that all event rarity comes from the auxiliary coordinate: $\rho_x=1$ and $\rho_u=C$. Then $\Gap(K_{\rm PM})=C\Gap(Q)$, so PM offers no advantage over rejection sampling. Both methods refresh $u$ independently and accept only when $u\in A_u$; local moves in $x$ cannot make this event more likely. Since $x$ and $u$ play analogous roles in the augmented target $\widetilde\pi$, one could also use local proposals for $u$, as in correlated PM (Remark~\ref{rmk:correlated_pm}).

In the selective-inference applications considered here, however, we expect selection rarity to arise mainly from the statistic $t$, particularly when external randomization is limited. This motivates our focus on Algorithm~\ref{alg:pm}, which refreshes $u$ independently. In the numerical experiments of Section~\ref{sec:experiments}, this algorithm substantially improves sampling efficiency over rejection sampling, while correlated PM provides a relatively modest additional improvement through local proposals for $u$.

For general events, Equation~\eqref{eq:exact-flow-comparison} gives the exact Dirichlet-form comparison
\begin{equation*}
 \calE_{K_{\rm PM}}(f,f)
 =\frac12\int\max\big\{L(x),L(y)\big\} \big(f(y)-f(x)\big)^2
       \pi(\rd x)K_{\rm or}(x,\rd y).
\end{equation*}
Taking the infimum over square-integrable functions with positive variance yields
\begin{equation*}
\Gap(K_{\rm PM})\geq\ell_-\Gap(K_{\rm or}),\qquad \ell_-:=\operatorname*{ess\,inf}_\pi L.
\end{equation*}
This bound is a specialization of the result of \citet{andrieu2015convergence} that pseudo-marginal samplers inherit a positive spectral gap from the corresponding marginal sampler when the weights are uniformly bounded. Writing $\gamma=\Gap(K_{\rm or})$, we obtain $\Gap(K_{\rm PM})\geq\gamma\ell_-$. Since rejection sampling has spectral gap $\Gap(K_{\rm R})=C$, the condition $\gamma\ell_->C$ guarantees that PM has a larger gap. When $\ell_->0$, this condition is equivalent to $\gamma>C/\ell_-$: the oracle gap $\gamma$ must exceed the rejection sampling gap $C$ by a factor greater than $1/\ell_-$. In the selective-inference applications of interest, including the file-drawer problem, we have $\ell_-=0$, so this bound is vacuous. This limitation motivates the weak Poincar\'e analysis in the next section.

\section{Sample complexity analysis}
\label{sec:general_theory}

\subsection{Weak Poincar\'e inequality}

When $L$ is not bounded away from zero, the preceding spectral-gap comparison becomes uninformative. We therefore identify a region carrying most of the selective probability where $L$ has a positive lower bound, and control the contribution of its complement. This localization yields finite-accuracy variance bounds for bounded observables even when the global spectral gap vanishes. It follows the weak Poincar\'e approach developed for PM samplers by \citet{andrieu2022comparison,andrieu2026weak}.

In selective inference, we use the ergodic average $\widebar f_N=\frac1N\sum_{i=1}^N f(X_i)$ to estimate expectations that determine selective $p$-values; see Section~\ref{sec:sampling-selective}. Since the Monte Carlo variance $\Var{\widebar f_N}$ affects the accuracy of the resulting inference, we measure sampling cost by the number of iterations needed to reduce this variance to a prescribed level. Specifically, suppose that $\Var[\pi]{f}<\infty$ and that $X_1,\ldots,X_N$ are generated by a $\pi$-reversible kernel $K$. For $\ep\in(0,1)$, define the Monte Carlo sample complexity as
\begin{equation}
 N_K(f;\varepsilon)=\inf\Bigl\{N\geq1:\,
     \Var{\widebar f_N}\leq\ep\Var[\pi]{f}\Bigr\}.
 \label{eq:mc-time}
\end{equation}
For the PM sampler in Algorithm~\ref{alg:pm}, each transition requires one query of the event indicator $\Indc{(x,u)\in A}$. Its sample complexity therefore also measures the number of queries needed to meet this variance criterion. We write $N_{\rm PM}=N_{K_{\rm PM}}$ for the sample complexity of the PM sampler.

To bound $N_{\rm PM}$, we compare the PM kernel $K_{\rm PM}$ in~\eqref{eq:marginal-kernel} with the oracle kernel $K_{\rm or}$ in~\eqref{eq:oracle-kernel}, both of which use the same $p$-reversible proposal kernel $Q$. Throughout this section, we assume that $Q$ is positive semidefinite, a condition used in the variance analysis below. This holds, for example, for random-walk Metropolis with Gaussian proposals~\citep{andrieu2015convergence}. We also assume that $f$ is essentially bounded, as is the case for the indicator functions typically used in selective inference. Define its essential oscillation under $\pi$ by $\osc(f)=\operatorname*{ess\,sup}_{\pi} f-\operatorname*{ess\,inf}_{\pi} f$.

\begin{theorem}[Localized sample complexity]
\label{thm:finite-accuracy}
Assume that $Q$ is $p$-reversible and positive semidefinite.
Let $H\subseteq \sfX$ be measurable, with $\pi(H)\geq1-\delta$ for some $\delta\in[0,1)$. Suppose that $L\geq\ell$ $\pi$-a.s. on $H$ for some $\ell\in(0,1]$ and that $\Gap(K_{\oracle,H})\geq\gamma$ for some $\gamma>0$, where $K_{\oracle, H}$ is the oracle kernel restricted to $H$. Let $X_1,\ldots,X_N$ be generated by $K_{\PM}$, with $X_1\sim \pi$. Then, for every bounded $f$, the variance of $\widebar f_N=\frac1N \sum_{i=1}^N f(X_i)$ satisfies
\begin{equation*}
 \Var{\widebar f_N}\leq
 \frac{2}{N\ell\gamma}\Var[\pi]{f}+\delta\osc(f)^2,
 \qquad N\geq1.
\end{equation*}
Moreover, if $\osc(f)^2\leq R \Var[\pi]{f}$ for some $R>0$, and $\delta\leq\varepsilon/(2R)$, then the sample complexity satisfies
\begin{equation}
 N_{\rm PM}(f;\ep)\leq\left\lceil\frac{4}{\varepsilon\ell\gamma}\right\rceil.
 \label{eq:general-cost}
\end{equation}
\end{theorem}
The proof is given in Appendix~\ref{sec:typical-region-proof}.
The theorem assumes that the chain is stationary. This assumption can be relaxed to an $M$-warm start, with $X_1\sim \nu$ and $\rd\nu / \rd\pi\leq M < \infty$ $\pi$-a.s. When $\delta\leq \ep/(2RM)$, the sample complexity bound in \eqref{eq:general-cost} incurs an additional factor of $M$; see Appendix~\ref{app:warm_start}.

The upper bound~\eqref{eq:general-cost} separates three sources of computational cost: $1/\ep$ captures the usual dependence on Monte Carlo accuracy, $1/\gamma$ reflects the cost of exploring $H$ with the censored oracle kernel $K_{\oracle,H}$, and $1/\ell$ accounts for the additional cost of estimating $L$ through auxiliary simulation rather than evaluating it exactly.

Rejection sampling requires $\lceil\ep^{-1}\rceil$ independent accepted draws to attain $\ep$-accuracy. Since each proposal is accepted with probability $C$, the expected number of proposals is
\begin{equation*}
 N_{\rm R}(\varepsilon)=\frac{\lceil\varepsilon^{-1}\rceil}{C},
\end{equation*}
which is also the expected number of queries to the selection event.
For fixed $\ep$ and $R$, the sample complexity bound for $K_{\PM}$ in \eqref{eq:general-cost} is much smaller than this rejection cost when
\[
 \gamma \gg\frac{C}{\ell},
 \qquad \delta\leq\frac{\ep}{2R}.
\]

To interpret this condition, suppose that $L(x)$ is within a multiplicative factor of $\ell$ for $p$-almost every $x\in H$. Specifically, suppose that $\ell\leq L(x)\leq c\ell$ for $p$-almost every $x\in H$, for some $c\geq1$, and that $\pi(H)=1-\delta$. Using \(C(1-\delta)=\int_H p(x)L(x)\,\rd x\), we obtain
\[
 \frac{p(H)}{1-\delta}
 \leq \frac{C}{\ell}
 \leq \frac{c\,p(H)}{1-\delta}.
\]
Therefore, $C/\ell$ differs from $p(H)$ by at most a fixed multiplicative factor.
The condition $\gamma\gg C/\ell$ can thus be read as $\gamma\gg p(H)$: the gap of the oracle kernel $K_{\oracle,H}$ is much larger than the probability that an independent prior proposal reaches $H$.
In the product-event example of Section~\ref{sec:product-event-comparison}, taking $H=A_x$ and $\ell=\rho_u$ gives $C/\ell=p(H)$ exactly; the argument above provides the analogous interpretation for general selection events.

This comparison highlights the key advantage of pseudo-marginal sampling over rejection sampling. When the event probability $L$ is large on a typical set $H$ and $Q$ explores $H$ efficiently, PM can retain states in this region across iterations. Rejection sampling, by contrast, draws afresh from the prior at every attempt, which is costly when $H$ is rare under the prior.

\subsection{The file drawer example}
\label{sec:file-drawer-analysis}

We now specialize the general comparison results to the file-drawer problem introduced in Section~\ref{sec:file_drawer}, where we obtain precise asymptotic results for the relative efficiency of PM and rejection sampling. Let $X$ and $U$ be independent standard normal variables with densities $p$ and $m$, respectively. The selection event is
\begin{equation*}
 A=\Big\{(x,u)\in\R^2:\, \sqrt a x+\sqrt{1-a}u>b \Big\},
\end{equation*}
where $a\in(0,1)$ and $b\in\mathbb{R}$. We set $\theta_0=0$ without loss of generality, since its effect can be absorbed into the threshold $b$ without changing the computational efficiency analysis. Increasing $b$ makes selection rarer, while increasing $a$ shifts more of the selection burden to $x$ and less to $u$.

We compare the Monte Carlo sample complexities of PM and rejection sampling as $b$ increases, allowing $a=a_b$ to vary with $b$. In Algorithm~\ref{alg:pm}, we use the random-walk Metropolis (RWM) kernel $Q$ with Gaussian proposal density $q(x,x')=h^{-1}\varphi((x'-x)/h)$ and acceptance probability $\alpha(x,x')=1\wedge p(x')/p(x)$. The resulting kernel is
\begin{equation}\label{eq:filedrawer-rwm}
Q(x,\rd x')=q(x,x')\alpha(x,x')\,\rd x'
+\left[1-\int q(x,y)\alpha(x,y)\,\rd y\right]\delta_x(\rd x').
\end{equation}
For the asymptotic analysis, we set $h=h_b=\lambda/(b\sqrt a)$ for a fixed $\lambda>0$, so that the acceptance probabilities remain nondegenerate as $b$ increases.

We use the subscript $b$ to indicate dependence on the threshold, writing $\pi_b$ for the target and $K_{\PM, b}$ for the corresponding PM kernel. We denote the PM sample complexity by $N_{\mathrm{PM},b}(f_b;\varepsilon):=N_{K_{\mathrm{PM},b}}(f_b;\varepsilon)$ and the rejection-sampling sample complexity by $N_{\mathrm{R},b}(f_b;\ep)$.

\begin{theorem}
\label{thm:filedrawer-cost}
Fix $a_0\in(0,1)$, $\lambda>0$, $\varepsilon\in(0,1)$, and $R>0$, and let $a=a_b\in[a_0,1)$ for each $b$. Suppose that the PM sampler uses the RWM kernel~\eqref{eq:filedrawer-rwm} with step size $h_b=\lambda/(b\sqrt{a_b})$ and that the chains start in stationarity. Let $f_b$ be any sequence of bounded test functions satisfying $\Var[\pi_b]{f_b}>0$ and $\osc(f_b)^2\leq R\Var[\pi_b]{f_b}$. Then, as $b\to\infty$,
\begin{align*}
 \log N_{\mathrm{R},b}(f_b ; \ep)&=\frac{b^2}{2}+\log b+O(1),\\
 \log N_{\mathrm{PM},b}(f_b; \ep)&=\frac{(1-a_b)b^2}{2}+O(1+b\sqrt{1-a_b}).
\end{align*}
In particular, if $b^2(1-a_b)=O(1)$, then $N_{\rm PM,b}(f_b; \ep)=O(1)$.
\end{theorem}
The proof in Appendix~\ref{sec:filedrawer-cost-proof} applies Theorem~\ref{thm:finite-accuracy} with a suitable choice of the typical region $H$ and an error level $\delta$. A conductance argument gives the required spectral gap bound for the restricted oracle kernel $K_{\oracle,H}$, using log-concavity of the target distribution.

By Theorem~\ref{thm:filedrawer-cost}, as $b\to\infty$,
\[
\log\frac{N_{\rm PM,b}(f_b; \ep)}{N_{\rm R,b}(f_b ; \ep)}
=-\frac{a_b b^2}{2} \big(1+o(1)\big).
\]
Thus, PM has exponentially smaller sample complexity than rejection sampling as $b$ increases. If $b\sqrt{1-a_b}=O(1)$, its sample complexity remains bounded even as selection becomes rarer. In this model, a randomization level $\sqrt{1-a_b}=O(1/b)$ is therefore sufficient for bounded PM cost, providing a computational consideration when choosing the amount of randomization.

Following Remark~\ref{rmk:correlated_pm}, we can replace the independent auxiliary proposal $u'\sim m$ with a correlated proposal, yielding the correlated pseudo-marginal (CPM) algorithm. Specifically, we use the Crank--Nicolson proposal kernel of \citet{deligiannidis2018correlated}:
\begin{equation}\label{eq:crank-nicolson}
 u'\sim R_\rho(u,\cdot)=\N(\rho u,1-\rho^2),\qquad 0\leq\rho<1.
\end{equation}
We accept the proposed pair $(x',u')$ if it belongs to $A$; otherwise, we retain the current values of both coordinates.

\begin{theorem}[File-drawer CPM complexity]
\label[theorem]{thm:filedrawer-cpm-cost}
Under the assumptions of Theorem~\ref{thm:filedrawer-cost}, fix $a_b=a\in[a_0,1)$ and $c>0$, and set $\rho_b=1-c/b^2$ for $b\geq\max\{1,\sqrt c\}$.
Let $N_{\mathrm{CPM},b}(f_b;\varepsilon)$ denote the sample complexity
in~\eqref{eq:mc-time}, computed from the $x$-coordinate of the stationary augmented
CPM chain. Then, as $b\to\infty$,
\[
 N_{\mathrm{CPM},b}(f_b;\varepsilon)=O(b^2).
\]
\end{theorem}
The proof is given in Appendix~\ref{sec:filedrawer-cpm-proof}. For fixed $a\in[a_0,1)$, independent PM has sample complexity exponential in $b^2$, whereas CPM has at most quadratic sample complexity. Correlating successive auxiliary proposals therefore avoids the exponential cost of independent refreshment, quantifying the advantage of CPM over independent PM in the file-drawer model.

\begin{figure}[H]
    \centering
    \includegraphics[width=0.8\linewidth]
    {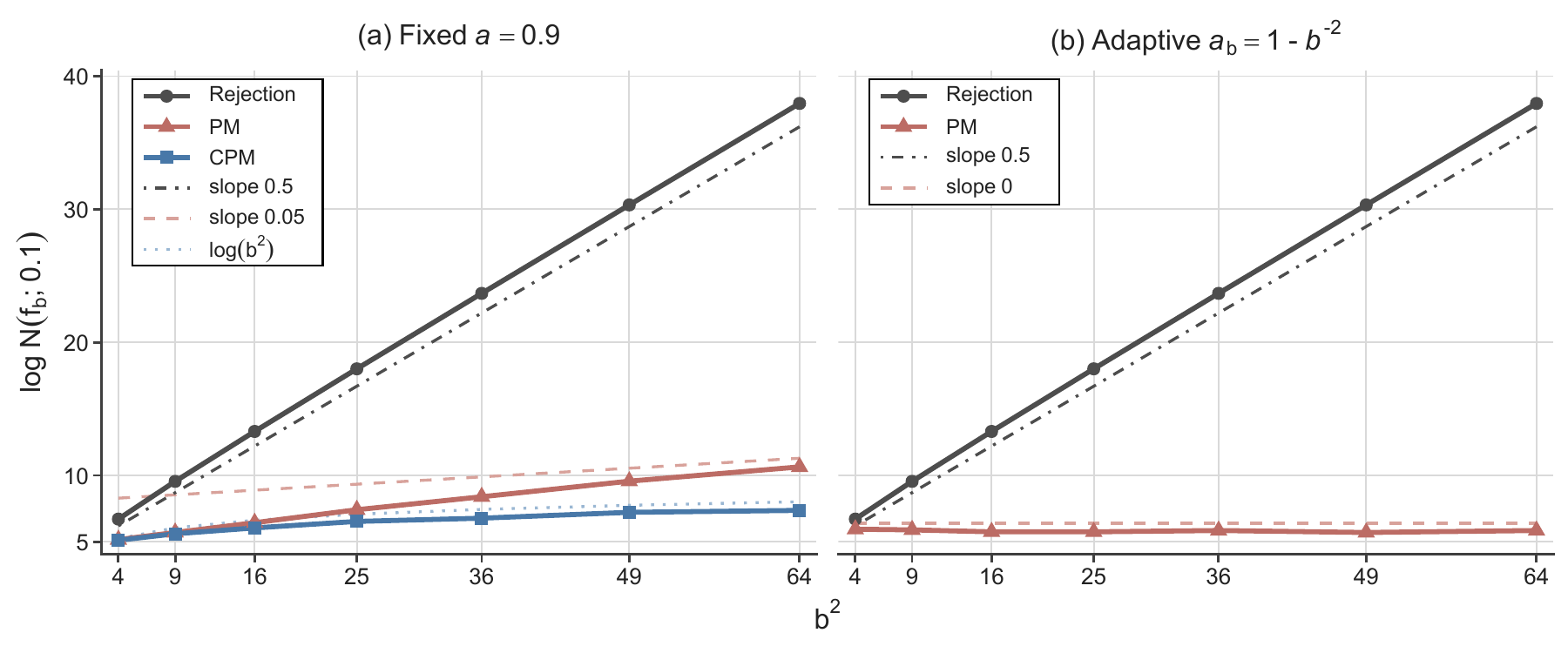}
    \caption{Sample complexity of rejection sampling, independent PM, and CPM in the file-drawer model. We plot $\log N(f_b;0.1)$ against $b^2$ for $f_b(x)=\mathbf{1}\{x\leq m_b\}$, where $m_b$ is the median of $\pi_b$. Panels (a) and (b) use $a_b=0.9$ and $a_b=1-b^{-2}$, respectively. Solid curves show the numerically computed sample complexities, and dashed curves show the theoretical scalings.}
    \label{fig:theorem43-verification}
\end{figure}

We numerically verify the sample-complexity scalings predicted by Theorems~\ref{thm:filedrawer-cost} and~\ref{thm:filedrawer-cpm-cost}. We use the observable $f_b(x)=\Indc{x\leq m_b}$, where $m_b$ is the median of $\pi_b$. We estimate the complexities of independent PM and CPM using $1{,}000$ simulation replicates and compute the rejection-sampling complexity analytically. Both PM methods use a Gaussian RWM proposal with step size $h_b=\lambda/(b\sqrt{a_b})$, where $\lambda=1$; for CPM, we set $\rho_b=1-c/b^{2}$ with $c=1$.
Figure~\ref{fig:theorem43-verification} plots the logarithms of the sample complexities against $b^2$.
The numerical results closely follow the predicted scalings, with both independent PM and CPM requiring substantially fewer selection-event queries than rejection sampling.

Our localized sample-complexity analysis builds on weak Poincar\'e comparisons that yield subgeometric convergence bounds for pseudo-marginal samplers in terms of their marginal counterparts and the tails of the estimator weights \citep{andrieu2022comparison,andrieu2026weak}. Earlier work studies how likelihood-estimation noise affects convergence and asymptotic variance relative to the ideal Metropolis--Hastings sampler \citep{andrieu2015convergence}, and how reducing this noise can improve efficiency for a fixed proposal \citep{andrieu2014establishing}. Related work provides theoretical guidance on tuning pseudo-marginal algorithms for computational efficiency, including the choice of estimation effort and proposal scale \citep{doucet2015efficient,sherlock2015efficiency,sherlock2016optimal,bornn2017use}. Our analysis compares PM directly with rejection sampling at a prescribed Monte Carlo accuracy, showing how efficient exploration of regions favored by selection can reduce cost as selection becomes rare.

\section{Numerical experiments}
\label{sec:experiments}

\subsection{Experiment setup}
\label{sec:experiment-setup}
We consider selective inference for two selection procedures with analytically intractable selection events: lasso variable selection with a cross-validated regularization parameter and changepoint detection with adaptive selection of the number of changepoints.
In both tasks, the observation $Y\in\R^n$ is modeled as $Y\sim\N_n(\mu_Y,\sigma^2 I_n)$. For the lasso, we test whether the coefficients of the best linear predictor of $Y$ using the selected covariates are zero. For changepoint detection, we test whether the mean shifts at the detected changepoints. As we will see, naive inference that does not account for the selection procedure is generally invalid.

For the observed selection outcome $M$, let $H_M\in\R^{d\times n}$ be the full-row-rank matrix defining the coefficients of the best linear predictor or the detected mean shifts, as specified below. We test $H_0:\theta_M=\mathbf{0}$, where $\theta_M=H_M\mu_Y$, using the statistic $T=H_MY\in\R^d$. For fixed $M$, the pre-selection distribution of $T$ is $\N_d(\theta_M,\Sigma_M)$, where $\Sigma_M=\sigma^2 H_M H_M^\top$.
We can decompose $Y$ as
\begin{equation}\label{eq:applications-reconstruction}
Y = \sigma^2 H_M^\top\Sigma_M^{-1} T + (I - \sigma^2 H_M^\top\Sigma_M^{-1} H_M) Y.
\end{equation}
For fixed $M$, $T$ is independent of $S=(I-\sigma^2 H_M^\top\Sigma_M^{-1}H_M)Y$ before conditioning on selection. Hence, conditioning on $S=s$ leaves the null pre-selection density of $T$ as $p_{\theta_0}(t)=\varphi(t;\mathbf{0},\Sigma_M)$.

Holding $S=s$ fixed, a candidate value $T=t$ determines the data vector $Y(t)=\sigma^2 H_M^\top\Sigma_M^{-1}t+s$ through the decomposition in \eqref{eq:applications-reconstruction}. We apply the selection procedure $\widehat M$ to $Y(t)$ with external randomization $W=\tau U$, where $U\sim\N_n(0,I_n)$ is independent of $Y$ and $\tau>0$ determines the randomization level. The probability of reproducing the observed selection outcome $M$ is
\begin{align*}
    L(t) = \PP{\widehat M(Y(t),W)=M},
\end{align*}
where the probability is over $W$. Conditional on $\widehat M=M$ and $S=s$, the selective density of $T$ is then proportional to $\varphi(t;\mathbf{0},\Sigma_M) L(t)$.

Because the selection probabilities are analytically intractable, we use the sampling approach of Section~\ref{sec:sampling-selective} to perform selective inference. We compare the pseudo-marginal sampler using independent auxiliary refreshment (PM, Algorithm~\ref{alg:pm}), its correlated variant with the Crank--Nicolson proposal~\eqref{eq:crank-nicolson} at $\rho=0.8$ (CPM, Remark~\ref{rmk:correlated_pm}), and rejection sampling that proposes from the pre-selection distribution. We initialize PM at the observed value of $T$ and CPM at the observed pair $(T,W)$, which reproduces the observed selection outcome. Additional experimental details and diagnostics appear in Appendix~\ref{app:additional-numerics}. Code to reproduce the experiments is available at \url{https://github.com/jialihe0927/pseudo-marginal-selinf}.

\subsection{Lasso with cross-validation}
\label{sec:applications-cvlasso}

We consider a Gaussian regression model with $n=200$ observations and $p=100$ covariates. Conditional on the design matrix $X\in\R^{n\times p}$, the response satisfies $Y\sim\N_n(X\beta,I_n)$.
We generate $X$ with independent rows from $\N(0,\Sigma_X)$, where $(\Sigma_X)_{jk}=0.5^{|j-k|}$, and then center and scale each column to have norm $\sqrt n$. The design matrix $X$ is fixed throughout. We set $\beta=\delta v$, where $v=(1,-0.8,0.6,-0.5,0.4,0,\ldots,0)\tran$ and $\delta\in\R$ controls the signal strength.

The selection procedure uses the perturbed response $Y+W$, where $W\sim\N_n(0,\tau^2I_n)$ is external randomization with $\tau=0.15$. For a given regularization parameter $\lambda>0$, the lasso solution is
\begin{align*}
    \hat\beta^\lambda
    =
    \underset{\beta\in\R^p}{\argmin}
    \left\{
    \frac{1}{2n}\|Y+W-X\beta\|_2^2
    +\lambda\|\beta\|_1
    \right\}.
\end{align*}
We choose $\hat\lambda$ by 10-fold cross-validation (CV) on $Y+W$, minimizing the cross-validated quadratic loss over a grid of 32 values. Fitting the lasso to the full perturbed response at $\hat\lambda$ gives the selected variable set
\begin{align*}
    \widehat M = \left\{j\in[p]:\, \hat\beta^{\hat\lambda}_{j} \neq 0 \right\}.
\end{align*}
Thus, $\widehat M$ includes both cross-validation to choose $\hat\lambda$ and the final lasso fit that selects the active variables.

For a nonempty selected set $M$ with $d=|M|$, let $X_M\in\R^{n\times d}$ be the submatrix of selected covariates and define $H_M=(X_M^\top X_M)^{-1}X_M^\top$. Then $T=H_MY$ is the least-squares coefficient vector for $X_M$. For fixed $M$, it has pre-selection distribution $\N_d(\theta_M,\Sigma_M)$, where $\theta_M=H_MX\beta$ is the coefficient vector of the best linear predictor of $Y$ based on $X_M$~\citep{berk2013valid,lee2016exact}.
Conditional on $S$ as described in Section~\ref{sec:experiment-setup}, we test $H_0:\theta_M=\mathbf{0}$.

Both PM and CPM use the Gaussian random-walk proposal $T'\sim\N(T,\frac{h^2}{d}\Sigma_M)$. We initialize the scale at $h=0.8$ and adapt it during warm-up to target an acceptance rate of $\alpha_{\mathrm{acc}}^\star=0.2$~\citep{andrieu2008tutorial}. We then hold the final scale $h_{\mathrm{final}}$ fixed during sampling.

\paragraph{Sampling efficiency}
We compare the three samplers on one dataset generated with $\delta=0.3$, for which the selected variable set is $M=\{1,7,15,30,81,100\}$. The resulting selective distribution is six-dimensional. Each PM and CPM chain uses 500 warm-up iterations followed by $N=19{,}500$ sampling iterations. We compute the effective sample size (ESS) of each coordinate from the sampling iterations and report the minimum:
\begin{align*}
    \min_{1\leq j\leq d} \mathrm{ESS}(T_j),\qquad \mathrm{ESS}(T_j) = \frac{N}{\tau_{\mathrm{IAT}}(T_j)},
\end{align*}
where \(\tau_{\mathrm{IAT}}(T_j)\) is the integrated autocorrelation time for coordinate \(T_j\). For rejection sampling, ESS is the number of accepted draws.
We measure sampling efficiency as ESS per $1{,}000$ selector evaluations:
\[
1000\cdot\frac{\mathrm{ESS}}{N_{\mathrm{query}}}.
\]
Here $N_{\mathrm{query}}$ counts selector evaluations, including those during warm-up for PM and CPM. Each evaluation reruns cross-validation and the final lasso fit, making these evaluations the main computational cost in this experiment.
Table~\ref{tab:cv-efficiency} reports this efficiency and its ratio to rejection-sampling efficiency (Gain). For PM and CPM, it also reports the chain acceptance rate (Acceptance), the fraction of proposals for which the selector does not return the observed set $M$ (Mismatch), and $h_{\mathrm{final}}$.

\begin{table}[ht]
\centering
\begin{tabular}{lrrrrr}
\toprule
Method & Efficiency & Gain & Acceptance & Mismatch
& $h_{\mathrm{final}}$ \\
\midrule
Rejection & 0.043 & $1.0\times$ & -- & -- & -- \\
PM & 2.904 & $67.5\times$ & 0.193 & 0.717 & 0.524 \\
CPM & 7.648 & $177.9\times$ & 0.220 & 0.638 & 0.718 \\
\bottomrule
\end{tabular}
\caption{Sampling efficiency for the six-dimensional cross-validated lasso example, measured as effective samples per $1{,}000$ selector evaluations. Gain is relative to rejection sampling.}
\label{tab:cv-efficiency}
\end{table}

Rejection sampling accepts only $129$ of $3\times 10^6$ proposals, giving an estimated overall selection probability of $4.3\times10^{-5}$ under the pre-selection distribution. As shown in Table~\ref{tab:cv-efficiency}, PM and CPM achieve \textbf{67.5-fold} and \textbf{177.9-fold} gains in efficiency, respectively, over rejection sampling.

The PM mismatch rate of approximately $0.7$ means that about $30\%$ of its proposals reproduce $M$, corresponding to an average selection probability of roughly $0.3$ over the proposed states. This is far higher than the overall selection probability $4.3\times 10^{-5}$ under pre-selection proposals. The contrast is consistent with the typical-region mechanism analyzed through the weak Poincar\'e inequality in Section~\ref{sec:general_theory}: PM can propose within regions where selection is relatively likely even though it is rare under the pre-selection distribution, helping explain the observed efficiency gain over rejection sampling.

\paragraph{Type I error under the global null}
We assess selective Type I error using data generated with $\delta=0$, so the underlying model has no true signal. We retain datasets only when the selected set $M$ is nonempty and contains at most 10 variables.
We use the quadratic test statistic $f(t)=t^\top\Sigma_M^{-1}t$ to compute the selective $p$-value defined in \eqref{eq:selective_p_value}.
We compare three methods:
\begin{itemize}
    \item \textbf{Proposed:} CPM sampling with a selection event that accounts for both cross-validation and lasso selection.
    \item \textbf{Fixed-$\lambda$:} CPM sampling that treats the observed $\hat\lambda$ as fixed, accounting for the final lasso selection but not the cross-validation step.
    \item \textbf{Naive:} inference based on the linear model $Y\sim \N(X_M\theta_M, I_n)$ without adjusting for selection.
\end{itemize}

\begin{figure}[ht]
\centering
\includegraphics[width=\linewidth]{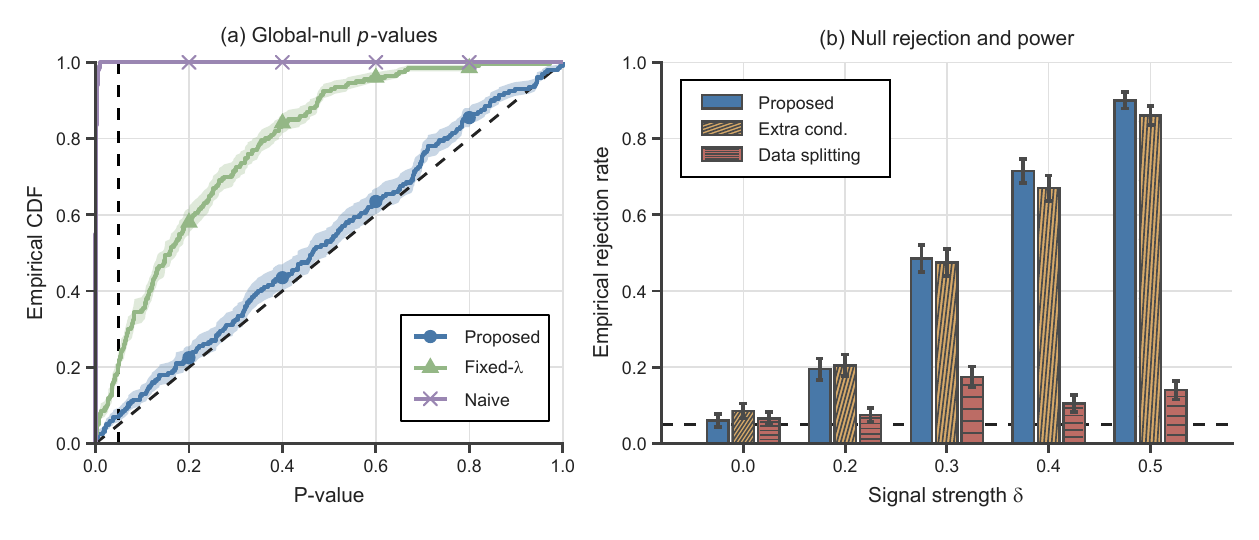}
\caption{Inference after cross-validated lasso. (a) Empirical CDFs of $p$-values under the global null; the diagonal dashed line is the uniform reference CDF, and the vertical dashed line marks $p=0.05$. (b) Empirical rejection rates at level $0.05$ across signal strengths. Shaded bands in (a) and error bars in (b) show pointwise standard errors across $200$ datasets.  }
\label{fig:cv-inference}
\end{figure}

For both CPM-based methods, we use $200$ warm-up iterations followed by $5{,}000$ sampling iterations and estimate selective $p$-values from the resulting samples using \eqref{eq:selective-pvalue-mc}. We repeat the experiment on 200 independent datasets. Figure~\ref{fig:cv-inference}(a) shows the empirical $p$-value distributions. The proposed method yields $p$-values close to uniform under the null, whereas the fixed-$\lambda$ and naive methods deviate substantially from uniformity. At the nominal level $0.05$, the fixed-$\lambda$ and naive methods reject at rates of about $0.2$ and $1$, respectively, indicating substantial Type I error inflation. These results suggest that accounting for the full selection procedure, including cross-validation to choose the regularization parameter, is important for Type I error control in this setting.

\paragraph{Power comparison}

We vary the signal strength $\delta$ over $\{0,0.2,0.3,0.4,0.5\}$ and compare empirical rejection rates, which measure Type I error at $\delta=0$ and power at positive signal strengths.
The proposed method conditions on the selected set $\widehat M=M$ without fixing the selected penalty or coefficient signs. We compare it with two alternatives:
\begin{itemize}
    \item \textbf{Extra conditioning:} also conditions on the selected penalty $\hat\lambda$ and the signs of the active lasso coefficients.
    \item \textbf{Data splitting:} uses $Y-\tau^{-2}W$ for inference, which is independent of the selection data $Y+W$. This amounts to conditioning on the entire selection data $Y+W$~\citep{fithian2014optimal} and is also known as data fission or thinning~\citep{garciarasines2023splitting,neufeld2024data,leiner2025datafission}.
\end{itemize}

The extra-conditioning method uses the same CPM sampler but accepts a proposal only if it reproduces the observed selected set $M$, penalty $\hat\lambda$, and active coefficient signs. By changing which selection outputs must match, the proposed sampler can target selective distributions tailored to different inferential questions.

For the proposed and extra-conditioning methods, we use CPM samples to estimate the selective $p$-value in \eqref{eq:selective-pvalue-mc} with $f(t)=a_M^\top\Sigma_M^{-1}t$, where $a_M=H_MXv$. For data splitting, we instead compute a one-sided $p$-value using $Y-\tau^{-2}W$, whose null distribution is $\N(0,(1+\tau^{-2})I_n)$.
Figure~\ref{fig:cv-inference}(b) reports empirical rejection rates at level $0.05$ across signal strengths $\delta$, with error bars showing one standard error across $200$ datasets.
At $\delta=0$, the extra-conditioning method has a slightly elevated empirical Type I error rate. Conditioning on the selected $\hat\lambda$ and signs of the active coefficients makes the event more restrictive and sampling more difficult: its median acceptance rate and tail-indicator ESS are $12\%$ and $51.4$, respectively, compared with $16\%$ and $102.4$ for the proposed method. The lower ESS suggests greater Monte Carlo error, which may contribute to the observed elevation.
At positive signal strengths, data splitting has substantially lower power than the proposed method, consistent with the variance inflation in $Y-\tau^{-2}W$ when $\tau=0.15$.
Extra conditioning has similar power to the proposed method at low signal strengths but lower power at higher strengths.

\subsection{Randomized changepoint detection}
\label{sec:applications-changepoint}

We consider a Gaussian sequence model with $n=200$ observations, $Y\sim\N_n(\mu_Y,I_n)$. For $i=1,\ldots,n$, the mean is $\mu_{Y,i}=\delta\,\mathbf1\{101\leq i\leq140\}$, where $\delta\in\R$ controls the signal strength. When $\delta\ne0$, the mean changes after observations 100 and 140.
Selection is based on the perturbed response $Y+W$, where $W\sim\N_n(0,\tau^2I_n)$ is independent of $Y$ and $\tau=0.1$.
We fit a piecewise-constant model to $Y+W$ using binary segmentation. For a fixed number $K$ of changepoints, the algorithm starts with the full sequence and makes $K$ splits. At each step, it chooses the split among the current segments that gives the largest reduction in residual sum of squares, subject to a minimum segment length of $10$. Let $M_k$ denote the changepoints after $k$ splits, with $M_0=\varnothing$.

We select the number of changepoints using the BIC-type criterion:
\[
\widehat K
=
\underset{0\leq k\leq4}{\argmin}\;
\left\{\mathrm{RSS}_k+1.25\,k\log n\right\},
\]
allowing for a maximum of four changepoints.
Here $\mathrm{RSS}_k$ is the residual sum of squares for the piecewise-constant fit to $Y+W$ defined by $M_k$, with $\mathrm{RSS}_0$ corresponding to the overall-mean model. The complete selection procedure includes both binary segmentation and the choice of $\widehat K$; its output is the final changepoint set $\widehat M=M_{\widehat K}$.
For binary segmentation with a fixed number of splits, \citet{hyun2021changepoint} show that fixing the ordered changepoint locations and their directions yields polyhedral selection events and, after suitable conditioning, truncated Gaussian tests. The polyhedral approach does not directly apply here, as it would require marginalizing over the added randomization~\citep{liu2023exact} and accounting for the adaptive choice of $K$.

For a nonempty changepoint set $M$ of size $d\geq1$, let $I_1,\ldots,I_{d+1}$ be its ordered segments. For $j=1,\ldots,d$, define $v_j=|I_{j+1}|^{-1}\mathbf1_{I_{j+1}}-|I_j|^{-1}\mathbf1_{I_j}$, where $\mathbf1_I\in\R^n$ is the indicator vector of segment $I$, and set $H_M=(v_1,\ldots,v_d)^\top$. Thus, $T=H_MY$ gives the differences between adjacent sample means, while $\theta_M=H_M\mu_Y$ gives the corresponding differences between segment-average means. Conditioning on $S$ as described in Section~\ref{sec:experiment-setup}, we test $H_0:\theta_M=\mathbf{0}$, which states that these segment averages are all equal.

\paragraph{Sampling efficiency}

We fix a dataset generated under $\delta=1.5$, with observed changepoint set $M=\{71,99,140\}$. The goal is to sample from the three-dimensional selective distribution. We use the same configurations and metrics as in Section~\ref{sec:applications-cvlasso} to compare the sampling efficiency. The results are summarized in Table~\ref{tab:cp-efficiency}, and the observations are similar to those in the previous example (Table~\ref{tab:cv-efficiency}). Independent rejection sampling produces $78$ accepted draws in $3\times 10^6$ proposals, giving an estimated selection probability of $2.60\times10^{-5}$. PM and CPM achieve \textbf{459.5-fold} and \textbf{594.3-fold} improvements in sampling efficiency over rejection sampling, respectively. This result further highlights the advantage of pseudo-marginal sampling when the selection is a rare event.

\begin{table}[ht]
\centering
\begin{tabular}{lrrrrr}
\toprule
Method  & Efficiency & Gain & Acceptance & Mismatch & $h_{\mathrm{final}}$ \\
\midrule
Rejection & 0.026 & $1.0\times$ & -- & -- & -- \\
PM & 11.948 & $459.5\times$ & 0.214 & 0.466 & 0.921 \\
CPM & 15.451 & $594.3\times$ & 0.184 & 0.432 & 1.206 \\
\bottomrule
\end{tabular}
\caption{Sampling efficiency for the three-dimensional
changepoint detection example.
Column definitions and efficiency measures follow
Table~\ref{tab:cv-efficiency}.}
\label{tab:cp-efficiency}
\end{table}

\paragraph{Type I error under the global null}

We simulate data under the global null $\delta=0$ and retain only datasets with a nonempty selected changepoint set. As in the Type I error experiment in Section~\ref{sec:applications-cvlasso}, we compare three methods: the proposed method, which correctly conditions on the full selection procedure; the fixed-$K$ method, which treats the observed $K$ as fixed; and the naive method, which ignores selection entirely. We use the same quadratic statistic for all three methods and the same CPM settings for the two selective methods, repeating the experiment on $200$ datasets.

\begin{figure}[ht]
\centering
\includegraphics[width=\linewidth]
{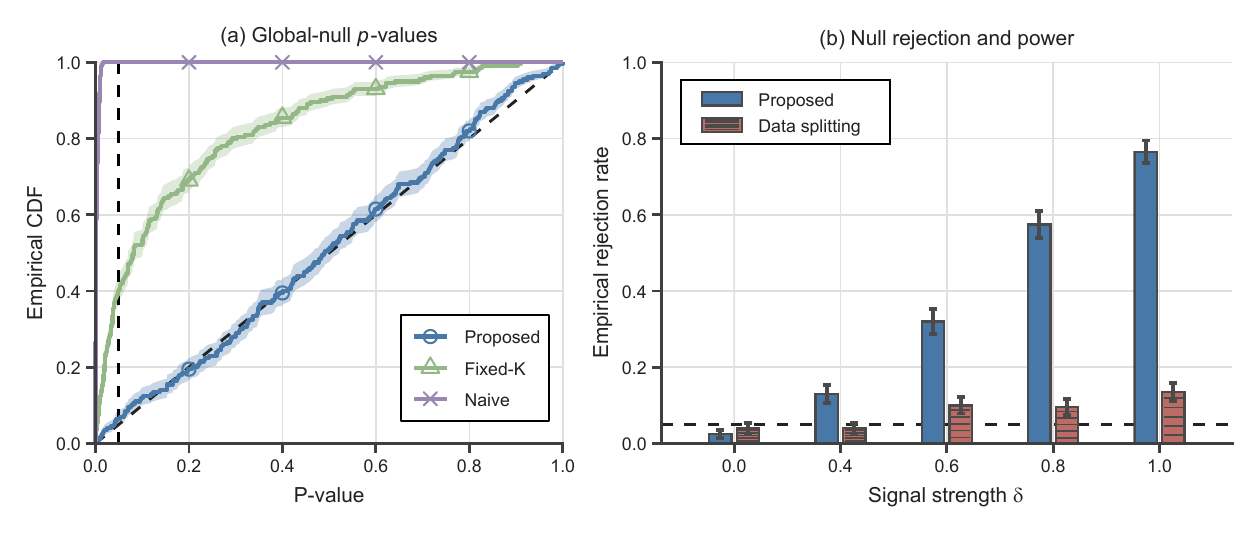}
\caption{Inference for randomized changepoint detection.
(a) Empirical $p$-value CDFs under the global null.
(b) Empirical rejection rates at level $0.05$ across signal strengths $\delta$.
Reference lines and error bars are as in Figure~\ref{fig:cv-inference}.}
\label{fig:cp-inference}
\end{figure}

Figure~\ref{fig:cp-inference}(a) shows the empirical $p$-value distributions. As in the lasso example, the proposed method yields $p$-values close to uniform. At the nominal level $0.05$, the fixed-$K$ method rejects about $40\%$ of null datasets, indicating substantial Type I error inflation when the adaptive choice of $K$ is ignored. The naive method shows even greater inflation.

\paragraph{Power comparison}
For signal strengths $\delta\in\{0,0.4,0.6,0.8,1.0\}$, we compare the proposed method with data splitting, as in the lasso power experiment in Section~\ref{sec:applications-cvlasso}. We take $a_M=H_M\nu$, where $\nu_i=\mathbf1\{101\leq i\leq140\}$, as the direction for the one-sided tests. The proposed method uses the CPM sampler described above and $f(t)=a_M^\top\Sigma_M^{-1}t$ to compute selective $p$-values. Data splitting uses $Y-\tau^{-2}W$ for the corresponding one-sided test.

At each signal strength, we analyze $200$ datasets with at least one selected changepoint. Figure~\ref{fig:cp-inference}(b) reports empirical rejection rates at level $0.05$. The proposed method has substantially higher power than data splitting at all four positive signal strengths,
while showing no Type I error inflation at $\delta=0$.

\section{Discussion}
\label{sec:discussion}

We proposed a pseudo-marginal MCMC sampler for selective distributions when the selection procedure can be simulated but its selection probability is unavailable analytically. The sampler can retain states for which selection is plausible instead of repeatedly drawing from the pre-selection distribution, as rejection sampling does. Our weak Poincar\'e analysis identifies conditions under which this advantage lowers Monte Carlo cost, even when the global spectral gap is zero. In the Gaussian file-drawer model, we quantified the sampling-efficiency gains of PM and correlated PM over rejection sampling. Numerical experiments also showed substantial gains and
illustrated the importance of accounting for hyperparameter selection in selective inference.

We have described how to test a single hypothesis using the pseudo-marginal sampler.
Constructing confidence sets requires tests at multiple null values $\theta$. Rather than rerunning MCMC for every null value, one can estimate expectations under $p_\theta(\cdot\mid\widehat M=M)$ by importance-weighting samples from $p_{\theta_0}(\cdot\mid\widehat M=M)$. Because $L(\cdot)$ does not depend on $\theta$, the weights are proportional to $p_\theta(t)/p_{\theta_0}(t)$, which is straightforward to compute.

Our current pseudo-marginal sampler uses a Gaussian random-walk proposal scaled by the pre-selection covariance, which need not match the shape of the selective distribution. Learning a proposal that better reflects this shape could improve sampling efficiency. Moreover, when the selection region is disconnected, local random-walk proposals may also struggle to move between components. Mixture proposals with components centered in different regions could help the chain traverse them~\citep{tjelmeland2001mode}. Finally, our current algorithm uses one auxiliary simulation ($B=1$) in every iteration. Adaptively allocating more simulations when needed may improve exploration of the tail regions of the selective distribution~\citep{abaoubida2025adaptive}.

\appendix

\section{Proofs for Section~\ref{sec:setup}}

\subsection{Proof of Theorem~\ref{thm:reversibility}}
\label{prf:reversibility}
\begin{proof}[Proof of Theorem~\ref{thm:reversibility}]
    Since $u'$ is refreshed independently in each iteration, the $x$-coordinate forms a Markov chain,
    with transition kernel
    \begin{align*}
        K_{\PM}(x,\rd x') &= Q(x, \rd x') \int m(\rd v) \indc{(x',v)\in A}
         + (1 - \int Q(x, \rd y) m(\rd v)\indc{(y,v)\in A})\delta_{x}(\rd x')\\
         &= Q(x, \rd x') L(x') + (1 - \int Q(x, \rd y) L(y))\delta_{x}(\rd x').
    \end{align*}
    For any measurable sets $B_1,B_2\subseteq\sfX$, the stationary flow is
    \begin{equation}\label{eq:transition_B1_B2}
    \begin{aligned}
    \int \indc{x\in B_1, x'\in B_2} \pi(\rd x) K_{\PM}(x,\rd x') &= \int\indc{x\in B_1, x'\in B_2} \frac{p(\rd x)L(x)}{C} Q(x, \rd x') L(x') \\
     &\qquad + \int_{B_1\cap B_2} \pi(\rd x) (1 - \int Q(x, \rd y)\,L(y)).
    \end{aligned}
    \end{equation}
    Since $Q$ is reversible with respect to $p$, the first term in~\eqref{eq:transition_B1_B2} can be written as
    \begin{align*}
        \frac1C \int\indc{x\in B_1, x'\in B_2} p(\rd x)Q(x, \rd x') L(x) L(x') &= \frac1C \int\indc{x\in B_1, x'\in B_2} p(\rd x')Q(x', \rd x) L(x) L(x')\\
        &=\frac1C \int\indc{x'\in B_1, x\in B_2} p(\rd x) Q(x, \rd x') L(x) L(x'),
    \end{align*}
    which is symmetric in $B_1$ and $B_2$. The second term in \eqref{eq:transition_B1_B2} is also symmetric in $B_1$ and $B_2$. The stationary flow in \eqref{eq:transition_B1_B2} is therefore symmetric in $B_1$ and $B_2$, establishing detailed balance with respect to $\pi$.
\end{proof}

\subsection{Proof of Remark~\ref{rmk:joint_pm_equivalence}}
\label{prf:joint_pm_equivalence}
\begin{proof}[Proof of Remark~\ref{rmk:joint_pm_equivalence}]
    Write $q(x,\rd y)=q(x,y)\,\rd y$ for the proposal kernel and its density. The usual PM algorithm retains a feasible joint state $(x,u)\in A$, proposes independently $x'\sim q(x,\cdot)$ and $u'\sim m$, and accepts the joint proposal with probability
    \[
        \alpha\bigl((x,u),(x',u')\bigr)
        =\min\left\{1,\frac{p(x')\widehat L(x', u')q(x',x)}{p(x)\widehat L(x, u)q(x,x')}\right\}
        =a_p(x,x')\indc{(x',u')\in A},
    \]
    where
    \[
        a_p(x,y)=\min\left\{1,\frac{p(y)q(y,x)}{p(x)q(x,y)}\right\},
    \]
    with the usual Metropolis--Hastings conventions when the denominator vanishes. Here we have used the facts that $\widehat L(x,u)=1$ at the current feasible state and that $\widehat L(x',u')$ is binary-valued. This factorization need not hold for a general nonbinary $\widehat L$.
    In particular, the acceptance probability does not depend on the retained $u$.
    Write $K_{\PM}^{(q)}$ for the $x$-marginal kernel of the usual PM formulation with proposal $q$.
    Integrating over $u'$ gives a Markov kernel for the $x$-coordinate:
    \[
        K_{\PM}^{(q)}(x,\rd x')
        =q(x,\rd x')a_p(x,x')L(x')
        +\left\{1-\int q(x,\rd y)a_p(x,y)L(y)\right\}\delta_x(\rd x').
    \]

    Now let $Q$ be the Metropolis--Hastings kernel targeting $p$ constructed from its proposal $q$:
    \begin{equation*}
        Q(x,\rd x')
        =q(x,\rd x')a_p(x,x')+h(x)\delta_x(\rd x'),
        \qquad
        h(x)=1-\int q(x,\rd y)a_p(x,y).
    \end{equation*}
    Substituting this expression into~\eqref{eq:marginal-kernel} yields
    \begin{align*}
        K_{\PM}(x,\rd x')
        &=q(x,\rd x')a_p(x,x')L(x')+h(x)L(x)\delta_x(\rd x')\\
        &\quad+\left\{1-\int q(x,\rd y)a_p(x,y)L(y)-h(x)L(x)\right\}\delta_x(\rd x')\\
        &=K_{\PM}^{(q)}(x,\rd x').
    \end{align*}
    Consequently, the two algorithms generate the same law for the $x$-process when started from the same $x$ with $L(x)>0$ and a feasible auxiliary state for the usual PM formulation, although their auxiliary-state updates may differ.
\end{proof}

\section{Proofs for Section~\ref{sec:general_theory}}
\label{sec:general-proof}

\subsection{Proof of Theorem~\ref{thm:finite-accuracy}}
\label{sec:typical-region-proof}

For a measurable set $H$ with $\pi(H)>0$, let $\pi_H(B)=\pi(B\cap H)/\pi(H)$.
For $x\in H$ and measurable $B\subseteq H$, define the censored oracle kernel by
\[
K_{\oracle,H}(x,B)
=K_{\oracle}(x,B\cap H)+K_{\oracle}(x,H^c)\indc{x\in B}.
\]
This kernel is reversible with respect to $\pi_H$.

\begin{proof}[Proof of Theorem~\ref{thm:finite-accuracy}]
We first derive a localized weak Poincar\'e inequality for $K_{\PM}$.
Fix a bounded function $g$ and let $c_H=\int g\,\rd\pi_H$. The variance of $g$ can be upper bounded by
\[
\Var[\pi]{g}
\leq
\int(g-c_H)^2\,\rd\pi
=
\int_H (g-c_H)^2\,\rd\pi
+
\int_{H^c} (g-c_H)^2\,\rd\pi.
\]
Note that $\Var[\pi_H]{g}=\int_H (g-c_H)^2\,\rd\pi_H =\frac{1}{\pi(H)}\int_H (g-c_H)^2\,\rd\pi$ and $|g-c_H|\leq\osc(g)$ $\pi$-a.s. Together with the assumption that $\pi(H^c)\leq \delta$, we have
\begin{align}\label{eq:var_g_bound}
      \Var[\pi]{g}\leq\pi(H)\Var[\pi_H]{g}+\delta\osc(g)^2.
\end{align}
From the exact Dirichlet-form comparison in Section~\ref{sec:product-event-comparison}, we have
\begin{align*}
 \calE_{K_{\rm PM}}(g,g)
 &\geq \frac12\int_{H\times H}\max\{L(x),L(y)\}\big(g(y)-g(x)\big)^2\pi(\rd x)K_{\rm or}(x,\rd y)\\
 &\geq \frac12 \ell \pi(H) \int_{H\times H} \big(g(y)-g(x)\big)^2\pi_H(\rd x)K_{\rm or}(x,\rd y)\\
 &=\ell\pi(H)\calE_{K_{{\rm or},H}}(g,g)
 \geq\ell\pi(H) \gamma \Var[\pi_H]{g}. \numberthis\label{eq:dirichlet_H_bound}
\end{align*}
Combining \eqref{eq:var_g_bound} and \eqref{eq:dirichlet_H_bound}, we obtain
\begin{equation*}
 \Var[\pi]{g}\leq
 \frac{\calE_{K_{\rm PM}}(g,g)}{\ell\gamma}
 +\delta\osc(g)^2.
\end{equation*}
Applying Lemma~\ref{lem:defective-mc} with $\kappa=\ell\gamma$ and $\beta=\delta$ yields the variance bound in Theorem~\ref{thm:finite-accuracy}. Note that Lemma~\ref{lem:defective-mc} requires that $K_{\PM}$ be positive semidefinite, which is guaranteed by the assumption that $Q$ is positive semidefinite. Indeed, for $h\in L^2(\pi)$, we have $Lh\in L^2(p)$ and
\begin{equation*}
 \langle h,K_{\rm PM}h\rangle_\pi
 =\frac1C\left\{\langle Lh,Q(Lh)\rangle_p
       +\int L(x)\big(1-QL(x)\big)h(x)^2\,p(\rd x)\right\}\geq0,
\end{equation*}
since $0\leq L\leq1$ implies $QL\leq1$.

Under the assumption that $\osc(f)^2\leq R \Var[\pi]{f}$, we choose $\delta\leq\varepsilon/(2R)$
and $n=\lceil4/(\varepsilon\ell\gamma)\rceil$. Then the variance bound just established implies
\begin{align*}
      \Var{\widebar{f}_n} \leq \frac{\ep}{2} \Var[\pi]{f} + \frac{\ep}{2} \Var[\pi]{f} = \ep \Var[\pi]{f},
\end{align*}
which proves $N_{\PM}(f;\ep)\leq \lceil4/(\varepsilon\ell\gamma)\rceil$.
\end{proof}

\begin{lemma}[Variance bound based on a weak Poincar\'e inequality]
\label[lemma]{lem:defective-mc}
Let $K$ be $\pi$-reversible and positive semidefinite.
Suppose for some $\kappa\in(0,2]$ and $\beta\geq0$,
\begin{equation}
 \Var[\pi]{g}\leq\kappa^{-1}\calE_K(g,g)+\beta\osc(g)^2
 \quad\text{for every bounded }g.
 \label{eq:defective-pi}
\end{equation}
Then for every bounded $f$, a stationary $K$-chain satisfies
\begin{equation*}
 \Var{\widebar f_n}\leq\frac{2}{n\kappa}\Var[\pi]{f}
                         +\beta\osc(f)^2,
 \qquad n\geq1.
\end{equation*}
\end{lemma}

\begin{proof}
The case with $n=1$ follows immediately from $\kappa\leq 2$; we assume $n\geq 2$ below.
Suppose $\mathbb{E}_{\pi}[f]=0$ without loss of generality. Denote $D=\osc(f)$ and $c_k=\langle f,K^k f\rangle_\pi$ for $k\geq 0$.
Because $K$ is positive semidefinite and self-adjoint,
\[
c_{k-1}-2c_k+c_{k+1}
=
\langle (I-K)f,\;K^{k-1}(I-K)f\rangle_\pi
\geq0,\qquad k\geq1.
\]
Since the chain is stationary, we have
\[
\Var{\widebar f_n}
=
\frac{1}{n^2}
\sum_{i=1}^n\sum_{j=1}^n
\Cov{f(X_i),f(X_j)}
=
\frac{1}{n^2}
\left\{
nc_0
+
2\sum_{k=1}^{n-1}(n-k)c_k
\right\}.
\]
We apply the bound $c_{2j+1}\leq(c_{2j}+c_{2j+2})/2$ to all the odd indices and obtain
\begin{align*}
\Var{\widebar f_n} &\leq \frac{1}{n^2}\Big\{
nc_0 + (n-1)(c_0 + c_2) + 2(n-2)c_2 + (n-3)(c_2+c_4) + 2(n-4)c_4+\ldots
\Big\}\\
&=\frac{1}{n^2}\Big\{(2n-1) c_0 + 4(n-2)c_2 + 4(n-4)c_4+\ldots \Big\}\\
&=\sum_{j=0}^{m}w_jc_{2j},\numberthis\label{eq:var-fn-bound-even}
\end{align*}
where $m=\lfloor n/2\rfloor$, and
\[
 w_j=\frac1{n^2}
 \begin{cases}
  2n-1, & j=0,\\
  4(n-2j), & 1\leq j<m,\\
  1, & j=m,\quad n=2m,\\
  4, & j=m,\quad n=2m+1.
 \end{cases}
\]
The coefficients $w_j$ satisfy
\begin{equation}\label{eq:wj-properties}
 \sum_{j=0}^{m}w_j=1,\qquad
 0\leq w_0\leq\frac2n,\qquad
 0\leq w_j\leq\frac4n\quad(1\leq j\leq m).
\end{equation}

Since $\mathbb{E}_{\pi}[K^jf]=0$ and $\osc(K^jf)\leq D$, applying the inequality~\eqref{eq:defective-pi} to $K^jf$ gives
\[
 c_{2j}=\Var[\pi]{K^jf}\leq\kappa^{-1}\calE_K(K^jf,K^jf)+\beta D^2.
\]
Denote $e_j=\calE_K(K^jf,K^jf)=c_{2j}-c_{2j+1}\geq0$.
Substituting this into \eqref{eq:var-fn-bound-even} and using \eqref{eq:wj-properties} gives
\begin{align*}
 \Var{\widebar f_n}&\leq \sum_{j=0}^m w_j(\kappa^{-1} e_j + \beta D^2 )
 =\kappa^{-1}\sum_{j=0}^{m}w_je_j+\beta D^2 \leq\frac{2}{n\kappa}
       \left(e_0+2\sum_{j=1}^{m}e_j\right)+\beta D^2.
\end{align*}
Since $c_{2j}\leq \frac{c_{2j-1}+c_{2j+1}}{2}$, we have $2e_j=2c_{2j}-2c_{2j+1} \leq c_{2j-1}-c_{2j+1}$. Telescoping gives
\[
 e_0+2\sum_{j=1}^{m}e_j
 \leq(c_0-c_1)+\sum_{j=1}^{m}(c_{2j-1}-c_{2j+1})
 =c_0-c_{2m+1}\leq \Var[\pi]{f}.
\]
This proves the desired inequality.
\end{proof}

\subsection{Warm-start mean-square bound}
\label{app:warm_start}

Let $\nu$ be an $M$-warm initial law, meaning $\rd\nu/\rd\pi\leq M<\infty$ $\pi$-a.s. Under the assumptions of Theorem~\ref{thm:finite-accuracy}, suppose $X_1\sim\nu$. If a bounded function $f$ satisfies $\osc(f)^2\leq R\Var[\pi]{f}$ for some $R>0$, then
\begin{equation*}
 \mathbb E_\nu\!\left[(\widebar f_N-\pi f)^2\right]
 \leq\varepsilon\Var[\pi]{f}
\end{equation*}
whenever $\delta\leq\varepsilon/(2MR)$ and $N\geq\lceil4M/(\varepsilon\ell\gamma)\rceil$, where $\pi f=\int f\,\rd\pi$.

\begin{proof}
Let $h=\rd\nu/\rd\pi\leq M$. Since the chains started from $\nu$ and $\pi$ use the same transition kernel, their path laws differ by the factor $h(X_1)$. Thus, by the variance bound in Theorem~\ref{thm:finite-accuracy} and stationarity under $\pi$,
\begin{align*}
 \mathbb E_\nu\!\left[(\widebar f_N-\pi f)^2\right]
 &=\mathbb E_\pi\!\left[h(X_1)(\widebar f_N-\pi f)^2\right]\\
 &\leq M\Var{\widebar f_N}\\
 &\leq \frac{2M}{N\ell\gamma}\Var[\pi]{f}
       +M\delta\osc(f)^2.
\end{align*}
Under the stated choices of $N$ and $\delta$, each term on the last line is at most $\varepsilon\Var[\pi]{f}/2$.
\end{proof}

\subsection{Proof of Theorem~\ref{thm:filedrawer-cost}}
\label{sec:filedrawer-cost-proof}

\begin{proof}[Proof of Theorem~\ref{thm:filedrawer-cost}]
Write $a=a_b$, $\theta=\theta_b:=b\sqrt{1-a_b}$, and
$q_\delta=\widebar\Phi^{-1}(\delta)$. Constants below are uniform over
$a\in[a_0,1)$ and sufficiently large $b$.
The Gaussian random-walk Metropolis kernel $Q_b$ is
$p$-reversible and positive semidefinite
\citep{andrieu2015convergence},
so the kernel assumption in Theorem~\ref{thm:finite-accuracy} is satisfied. In the proof below, we add subscript $b$ to all relevant quantities to emphasize their dependence on $b$.

The sample complexity of rejection sampling is $N_{\mathrm{R},b}(f_b;\varepsilon)=\lceil\varepsilon^{-1}\rceil/\widebar\Phi(b)$.
By Mills' ratio, we have $\widebar\Phi(b)^{-1}=\sqrt{2\pi}\,b e^{b^2/2}\{1+O(b^{-2})\}$. Therefore,
\begin{align*}
      \log N_{\mathrm{R},b}(f_b;\varepsilon)&=\frac{b^2}{2}+\log b+O(1).
\end{align*}

It remains to bound the sample complexity of the PM sampler. We will apply Theorem~\ref{thm:finite-accuracy} with a suitable typical set $H_{\delta,b}$ with probability at least $1-\delta$, and then show that the likelihood $L$ is bounded away from 0 on this set, and that the oracle kernel restricted to this set has a positive spectral gap.

\paragraph{Reparametrization of variables. }
We first introduce a reparametrization of the variables, which makes the analysis convenient.
Under the unconditional law of $(X,U)$, define
\[
 S=\sqrt a\,X+\sqrt{1-a}\,U,\qquad
 T=\sqrt{1-a}\,X-\sqrt a\,U.
\]
This is an orthogonal transformation of two independent standard
Gaussians, so $S$ and $T$ are also independent standard Gaussians.
Since $A_b=\{S>b\}$, conditioning on $A_b$ truncates only $S$, while
$T$ remains standard Gaussian and independent of $S$.
Inverting the transformation therefore gives the representation
\begin{equation}\label{eq:X-representation}
 X=\sqrt a\,S+\sqrt{1-a}\,T\sim\pi_b, \;\;\text{ where }\;\;
 S\sim\N(0,1)\mid\{S>b\},\quad T\sim\N(0,1),
\end{equation}
with $S,T$ independent.

We then rescale $X$ to
\[
 V=b\sqrt a(X-\sqrt a\,b).
\]
The inverse transformation is $x=\sqrt a\,b+v/(b\sqrt a)$, with
Jacobian $\rd x/\rd v=1/(b\sqrt a)$.
Since the target distribution is
\begin{align*}
\pi_b(x)=\frac{\varphi(x)\widebar\Phi\Big(\frac{b-\sqrt a\,x}{\sqrt{1-a}}\Big)}{C_b},\qquad C_b=\widebar \Phi(b),
\end{align*}
by the change of variables, the density of $V$ is
\[
 p_V(v)=\frac{\varphi(\sqrt a\,b+v/(b\sqrt a))}{b\sqrt a\,C_b}
        \widebar\Phi(\theta-v/\theta)
       =Z_V^{-1}F(v)G(v),\qquad v\in\mathbb R,
\]
where
\[
 F(v)=e^{-v-v^2/(2ab^2)},\qquad
 G(v)=\widebar\Phi(\theta-v/\theta),\qquad
 Z_V=\int_{\mathbb R}F(v)G(v)\,\rd v.
\]

Note that the target $\pi=\pi_b$ is log-concave, because the Gaussian density $\varphi$ is log-concave and the survival function $\widebar\Phi$ is log-concave as well. Since $V$ is an affine transformation of $X$, the density $p_V$ is also log-concave.

\paragraph{Typical set. }For fixed $\delta\in(0,1/2)$, let
\[
 H_{\delta,b}=\{x\geq\sqrt a\,b-\sqrt{1-a}\,q_\delta\}.
\]
From the representation of $X$ in \eqref{eq:X-representation}, $T\geq-q_\delta$ implies $X\in H_{\delta,b}$, which implies $\pi_b(H_{\delta,b})\geq1-\delta$.
Since $L_b(x)=\widebar\Phi(\frac{b-\sqrt a\,x}{\sqrt{1-a}})$ is increasing in $x$,
on $H_{\delta,b}$ we have
\[
L_b(x)
\geq
L_b(\sqrt a\,b-\sqrt{1-a}\,q_\delta)
=
\widebar\Phi(b\sqrt{1-a}+\sqrt a\,q_\delta)
=:
\ell_{\delta,b}.
\]

In the $V$-space, the typical set corresponds to
$J=[-\sqrt a\,\theta q_\delta,\infty)$. The density of $V$ restricted to $J$ is
\[
 \mu(v)=Z_J^{-1}F(v)G(v)\indc{v\in J},\qquad
 Z_J=\int_J F(v)G(v)\,\rd v.
\]

\paragraph{Variance bound of the truncated target. } We now prove that the variance of $\mu$, the law of $V$ restricted to the typical set $J$, is upper bounded by $\frac{2+\theta^2}{1-\delta} $.
First, note that
\begin{equation*}
 \sigma^2:=\Var{V\mid V\in J}\leq
 \mathbb E[(V-\mathbb EV)^2\mid V\in J]
 \leq\frac{\Var{V}}{\PP{V\in J}}\leq \frac{\Var{V}}{1-\delta}.
\end{equation*}
Substituting the representation of $X$ \eqref{eq:X-representation} into $V$, we have
\[
 V=b\sqrt a(X-\sqrt a\,b)=b\sqrt{a}(\sqrt{a} S - \sqrt{a}b + \sqrt{1-a} T ) =aE_b+\sqrt a\,\theta T,
 \qquad E_b=b(S-b).
\]
Since $S\sim \N(0,1)\vert_{[b,\infty)}$,
\begin{equation}\label{eq:filedrawer:excess-tail}
 \PP{E_b>u}=\frac{\int_{b+u/b}^\infty\varphi(s)\,\rd s}
        {\int_b^\infty\varphi(s)\,\rd s}
 =\frac{\int_b^\infty\varphi(s+u/b)\,\rd s}
        {\int_b^\infty\varphi(s)\,\rd s}
 \leq e^{-u},\qquad u\geq0,
\end{equation}
where the inequality is due to $\frac{\varphi(s+u/b)}{\varphi(s)}=e^{-su/b-u^2/(2b^2)}\leq e^{-u}$ for $s\geq b$ and $u\geq0$.
Hence, $E_b$ is stochastically smaller than $\mathrm{Exp}(1)$, and thus
$\mathbb E E_b^2  \leq2$.

Since $E_b$ and $T$ are independent, $\Var{V}=a^2\Var{E_b}+a\theta^2\leq 2a^2+a\theta^2\leq 2+\theta^2$.
This proves that
\begin{align}\label{eq:sigmasq_bound}
    \sigma^2 \leq \frac{2+\theta^2}{1-\delta}.
\end{align}

\paragraph{Oracle gap on the typical set}

In the rescaled coordinate in $V$, the Gaussian random-walk proposal is
\[
 v'=v+\lambda Z,\qquad Z\sim\N(0,1),
\]
which is accepted with probability
\begin{align*}
    1 \wedge \frac{\varphi(v'; - ab^2, ab^2)}{\varphi(v; -ab^2, ab^2)}=1\wedge \frac{e^{-v'-v'^2/(2ab^2)} }{e^{-v-v^2/(2ab^2)}}=1\wedge \frac{F(v')}{F(v)}.
\end{align*}
The oracle chain truncated to the typical set $J$ then accepts the state $v'$ with probability
\begin{align*}
    1\wedge \frac{L(v')}{L(v)} = 1\wedge \frac{\widebar\Phi(\theta - v'/\theta) }{\widebar\Phi(\theta-v/\theta )}=1\wedge \frac{G(v')}{G(v)}.
\end{align*}
Thus, the off-diagonal transition density of the oracle kernel restricted to $J$ is
\begin{equation}\label{eq:filedrawer:oracle-kernel}
 k(v,v')=q_\lambda(v'-v)
 \min\{1,F(v')/F(v)\}\min\{1,G(v')/G(v)\},\qquad v,v'\in J,
\end{equation}
where $q_\lambda$ is the $\N(0,\lambda^2)$ density; the remaining
probability is assigned to the current state. Denote this kernel by $K$.

We first show that the transition kernels from nearby states have substantial overlap.
\begin{lemma}[Overlap of nearby transitions]
\label[lemma]{lem:filedrawer:overlap}
Set $d_0=\frac14$, $D=1+\frac3{a_0}$, $\beta=q_\lambda(1/2)e^{-D/2}$, and $\alpha=\frac\beta4$.
For $b\geq1$ and all $v_1,v_2\in J$ with $|v_1-v_2|\leq d_0$,
\begin{equation}\label{eq:filedrawer:local-overlap}
 \int_J\min\{k(v_1,z),k(v_2,z)\}\,\rd z\geq\alpha>0.
\end{equation}
\end{lemma}

We now use a standard conductance argument~\citep[proof of Theorem~3.2]{lovasz1993random} to bound the spectral gap of $K$, which is reversible with respect to $\mu$.

Since $\mu$ is a one-dimensional log-concave law of variance $\sigma^2$,
\citet[Proposition~4.1]{bobkov1999isoperimetric} bounds its Cheeger
isoperimetric constant below by $1/(\sqrt3\,\sigma)$. Thus, for every
measurable $B\subset J$,
\[
 \mu^+(B)\geq\frac1{\sqrt3\,\sigma}
                  \min\{\mu(B),1-\mu(B)\},
\]
where $B^t=\{v\in J:\operatorname{dist}(v,B)<t\}$ and
$\mu^+(B)=\liminf_{t\downarrow0}\{\mu(B^t)-\mu(B)\}/t$ is the
outer boundary measure. This implies the following separator inequality:
\begin{equation}
 \mu\{J\setminus(B_1\cup B_2)\}
 \geq\frac{d_0}{\sqrt3\,\sigma}\min\{\mu(B_1),\mu(B_2)\},
 \qquad \operatorname{dist}(B_1,B_2)\geq d_0>0,
 \label{eq:cheeger_ineq}
\end{equation}
for measurable $B_1,B_2\subset J$.
Indeed, set $s=\min\{\mu(B_1),\mu(B_2)\}$. For $0<t<d_0$,
$B_1\subset B_1^t\subset J\setminus B_2$, so
$\min\{\mu(B_1^t),1-\mu(B_1^t)\}\geq s$.
Applying the isoperimetric bound to these enlargements gives
\[
 \frac{\rd}{\rd t}\mu(B_1^t)\geq\mu^+(B_1^t)
 \geq\frac{s}{\sqrt3\,\sigma}
 \qquad\text{for almost every }t\in(0,d_0).
\]
Integrating yields
$\mu(B_1^{d_0})-\mu(B_1)\geq d_0s/(\sqrt3\,\sigma)$.
Since $B_1^{d_0}\setminus B_1\subset J\setminus(B_1\cup B_2)$,
this proves \eqref{eq:cheeger_ineq}.

Fix $A\subset J$ with $\mu(A)\in(0,1/2]$.  Define
\[
 \mathcal Q(A,A^c)=\int_A K(x,A^c)\,\mu(\rd x),\qquad
 \Phi_K=\inf_{0<\mu(A)\leq1/2}
             \frac{\mathcal Q(A,A^c)}{\mu(A)}.
\]
We want to lower bound the conductance $\Phi_K$.
Set $\alpha$ as in Lemma~\ref{lem:filedrawer:overlap}, and let
\[
 A_0=\{x\in A:K(x,A^c)<\alpha/4\},\qquad
 B_0=\{y\in A^c:K(y,A)<\alpha/4\}.
\]
Outside these sets, each point crosses with probability at least
$\alpha/4$. If $\mu(A_0)<\mu(A)/2$, then
\[
 \mathcal Q(A,A^c)\geq\frac\alpha4\mu(A\setminus A_0)
 \geq\frac{\alpha }{8} \mu(A).
\]
Likewise, if $\mu(B_0)<\mu(A^c)/2$, reversibility gives
$\mathcal Q(A,A^c) \geq\alpha \mu(A^c)/8\geq\alpha \mu(A)/8$.
Hence, if either $\mu(A_0)<\mu(A)/2$ or $\mu(B_0)<\mu(A^c)/2$, we have
\begin{align*}
    \frac{\mathcal Q(A,A^c)}{\mu(A)}
    \geq \frac{\alpha}{8}.
\end{align*}
It remains to consider the case where $\mu(A_0)\geq \mu(A)/2$ and $\mu(B_0)\geq\mu(A^c)/2$.

For $v_1\in A_0$ and $v_2\in B_0$, we have
\begin{align*}
 \int_J\min\{k(v_1,z),k(v_2,z)\}\,\rd z
 &\leq\int_A k(v_2,z)\,\rd z+\int_{A^c}k(v_1,z)\,\rd z\\
 &= K(v_2,A)+K(v_1,A^c)<\frac\alpha2.
\end{align*}
If $|v_1-v_2|\leq d_0$, this contradicts the overlap lower bound $\alpha$
in Lemma~\ref{lem:filedrawer:overlap}. Thus $A_0$ and $B_0$ have distance
at least $d_0$.

Since $\mu(A_0)\geq \mu(A)/2$ and $\mu(B_0)\geq \mu(A)/2$, applying the separator inequality~\eqref{eq:cheeger_ineq} gives
\[
 \mu\bigl(J\setminus(A_0\cup B_0)\bigr)
 \geq\frac{d_0}{2\sqrt{3}\sigma}\mu(A) .
\]
By reversibility, we have
\begin{align*}
 2\mathcal Q(A,A^c)
 &=\mathcal Q(A,A^c)+\mathcal Q(A^c,A)\\
 &=\int_A K(x,A^c)\,\mu(\rd x)
   +\int_{A^c}K(y,A)\,\mu(\rd y)\\
 &\geq\frac\alpha4
       \bigl\{\mu(A\setminus A_0)+\mu(A^c\setminus B_0)\bigr\}
       =\frac\alpha4\mu\bigl(J\setminus(A_0\cup B_0)\bigr)\\
 &\geq\frac{\alpha d_0\mu(A)}{8\sqrt3\,\sigma}.
\end{align*}
Taking the infimum over $A$, we obtain
\[
 \Phi_K=\inf_{0<\mu(A)\leq1/2}\frac{\mathcal Q(A,A^c)}{\mu(A)}
 \geq\min\left\{\frac{\alpha}{8},
               \frac{\alpha d_0}{16\sqrt3\,\sigma}\right\}.
\]
Cheeger's inequality for reversible Markov kernels
\citep[Theorem~3.5]{lawler1988bounds} and the upper bound on $\sigma^2$ \eqref{eq:sigmasq_bound} imply
\[
 \Gap(K)\geq\frac{\Phi_K^2}{2}
 \gtrsim\frac1{1+\sigma^2}\gtrsim\frac1{1+\theta^2},
\]
where the implied constants depend only on $\lambda$ and $a_0$, uniformly over $\delta\in(0,1/2)$, $a\in[a_0,1)$, and $b\geq1$.

\paragraph{Putting everything together}

We have established that $\pi_b(H_{\delta,b})\geq1-\delta$, $L_b(x)\geq \ell_{\delta,b}$ on $H_{\delta,b}$, and that the oracle kernel restricted to $H_{\delta,b}$ has a spectral gap lower bound $\gamma_H=c/(1+\theta^2)$ for some constant $c$ depending only on $\lambda$ and $a_0$.
Theorem~\ref{thm:finite-accuracy} with $\delta=\ep/(2R)$ then yields
\begin{equation}\label{eq:filedrawer-explicit-upper}
 N_{\mathrm{PM},b}(f_b;\varepsilon)\leq
 \left\lceil\frac{4(1+\theta^2)}
 {c\varepsilon\widebar\Phi(\theta+\sqrt a\,q_\delta)}
 \right\rceil.
\end{equation}
Recall that $\theta=b\sqrt{1-a_b}$.
If $\theta$ is bounded, the right-hand side of \eqref{eq:filedrawer-explicit-upper} is bounded, and thus $N_{\mathrm{PM},b}(f_b;\varepsilon)=O(1)$.
As $\theta\to\infty$, Mills' ratio bounds the logarithm of the right-hand
side by $\theta^2/2+O\{\theta+\log(1+\theta)\}$.
A matching lower bound is proved in Lemma~\ref{lem:filedrawer:time-lower}. This proves $\log N_{\mathrm{PM},b}(f_b;\varepsilon)=\theta^2/2+O(1+\theta)$.

\end{proof}

\begin{proof}[Proof of Lemma~\ref{lem:filedrawer:overlap}]
Assume $v_1\leq v_2$ without loss of generality.
If $v_2\leq2$, consider the interval $I=[v_2,v_2+1/4]\subset J$. Since $G$ is increasing, the factor $\min\{1,G(z)/G(v_j)\}$ equals one for any $z\in I$ and $j\in\{1,2\}$.
For any $v\in(v_1, z) $, $v\leq 2+\frac14<3 $, and thus
\[
 (\log F)'(v)=-1-\frac{v}{ab^2}\geq -1 - \frac{3}{ab^2} \geq-D,
\]
since $a\geq a_0$ and $b\geq1$ by assumption.
Consequently $F(z)/F(v_j)\geq e^{-D(z-v_j)}\geq e^{-D/2}$ for $j\in\{1,2\}$. In addition, the distance between $v_j$ and $z$ is at most $1/2$, and thus $q_\lambda(z-v_j)\geq q_\lambda(1/2)$. Using the definition of $k(v,v')$ in \eqref{eq:filedrawer:oracle-kernel}, $k(v_j,z)\geq q_\lambda(1/2) e^{-D/2} = \beta$.

If $v_2>2$, consider the interval $I=[v_1-1/4,v_1]$.
Since $v_1\geq v_2-1/4>7/4$, this interval lies in $(1,\infty)\subset J$.
Here $F$ decreases, so the acceptance factor $\min\{1, F(z)/F(v_j)\}$ equals one for any $z\in I$ and $j\in\{1,2\}$.

To bound the other acceptance factor $\min\{1, G(z)/G(v_j)\}$, we will prove that $(\log G)'(v) <1$ for $v\in[z,v_2]$. To see this, let
$r(t)=\varphi(t)/\widebar\Phi(t)$. For $Z\sim\N(0,1)$,
integration by parts gives
\[
 0<\Var{Z\mid Z>t}=1+t\,r(t)-r(t)^2,
 \qquad\text{so}\qquad
 r(t)<\frac{t+\sqrt{t^2+4}}2=:r_{\rm ub}(t).
\]
Since $G(v)=\widebar\Phi(\theta-v/\theta)$,
\[
 (\log G)'(v)=\frac{\varphi(\theta-v/\theta)}{\widebar\Phi(\theta-v/\theta) \theta}
 =\frac{r(\theta - v/\theta)}{\theta}<\frac{r_{\rm ub}(\theta - v/\theta)}{\theta}.
\]
For $v\in[z,v_2]$, we have $v>1$ and hence
$r_{\rm ub}(\theta-v/\theta)<r_{\rm ub}(\theta-1/\theta)= \theta$. Thus, $(\log G)'(v)<1$, and
\[
 \log\frac{G(z)}{G(v_j)}
 =-\int_z^{v_j}(\log G)'(v)\,\rd v
 \geq-(v_j-z)\geq-\frac12.
\]
Thus $k(v_j,z)\geq q_\lambda(1/2)e^{-1/2}\geq\beta$ for $j\in\{1,2\}$ in this case too.
Integrating the common lower bound over an interval of length $1/4$
proves \eqref{eq:filedrawer:local-overlap}.
\end{proof}

\begin{lemma}[Matching lower bound]
\label[lemma]{lem:filedrawer:time-lower}
For the stationary PM chain and observables in
Theorem~\ref{thm:filedrawer-cost}, as $\theta\to\infty$,
\[
 \log N_{\mathrm{PM},b}(f_b;\varepsilon)
 \geq\frac{\theta^2}{2}-O\{\theta+\log(1+\theta)\}.
\]
The implied constant depends only on $a_0,\lambda,\varepsilon,R$,
uniformly over all admissible $f_b$ and $a\in[a_0,1)$.
\end{lemma}

\begin{proof}[Proof of Lemma~\ref{lem:filedrawer:time-lower}]
Until the chain first moves, its sample average equals its initial
observation. We show that a move is unlikely on the time scale below,
so averaging cannot yet reduce the variance to the prescribed level.

In this proof, take $\delta=(1-\varepsilon)/(4R)$ and use the quantile
notation $q_\delta=\widebar\Phi^{-1}(\delta)$ introduced above.
The representation~\eqref{eq:X-representation} gives
$X=\sqrt a\,b+\frac{\sqrt a}{b}E_b+\sqrt{1-a}\,T$.
Define an upper cutoff
\[
 x_+=\sqrt a\,b+\sqrt{1-a}\,q_{\delta/2}
      +\frac{\sqrt a}{b}\log(2/\delta).
\]
By the previously established tail bound~\eqref{eq:filedrawer:excess-tail}
and a union bound,
\[
 \pi_b\{X>x_+\}
 \leq\PP{T>q_{\delta/2}}+\PP{E_b>\log(2/\delta)}
 \leq\delta.
\]

Let $p_{\rm move}(x)$ be the probability that one PM transition changes
the current state:
\[
 p_{\rm move}(x):=\PP{X_{t+1}\ne X_t\mid X_t=x}
 =K_{{\rm PM},b}(x,\mathbb R\setminus\{x\}).
\]
For a Gaussian proposal $x+h_bZ$, where
$h_b=\lambda/(\sqrt a b)$ and $Z\sim\N(0,1)$, a move requires both
acceptance by the base Metropolis kernel targeting $\varphi$ and
success of the fresh auxiliary selection test. The corresponding
probabilities are $1\wedge\{\varphi(x+h_bZ)/\varphi(x)\}$ and
$L_b(x+h_bZ)$. Replacing the first factor by its upper bound one gives
\begin{equation}\label{eq:filedrawer:move-bound}
 \begin{aligned}
 p_{\rm move}(x)
 &=\mathbb E_Z\!\left[
    \left(1\wedge\frac{\varphi(x+h_bZ)}{\varphi(x)}\right)
    L_b(x+h_bZ)\right]\\
 &\leq\mathbb E_Z L_b(x+h_b Z)
 =\widebar\Phi\!\left(
    \frac{b-\sqrt a\,x}{\sqrt{1-a+ah_b^2}}\right).
 \end{aligned}
\end{equation}
The last equality follows by integrating over the independent standard
Gaussian proposal and auxiliary variables. This upper bound increases
in $x$, so for every $x\leq x_+$,
\[
 p_{\rm move}(x)\leq M_b
 :=\widebar\Phi\!\left(
 \frac{\theta^2-\sqrt a\,\theta q_{\delta/2}-a\log(2/\delta)}
      {\sqrt{\theta^2+\lambda^2}}\right).
\]

Let $D_n$ be the event of at least one move between the first $n$
observations. Conditional on $X_1\leq x_+$, the chain stays at $X_1$
until its first move, with move probability at most $M_b$ at each step.
Using $1-(1-p_{\rm move}(x))^{n-1}\leq (n-1)p_{\rm move}(x)$, we have
\[
 \PP{D_n}\leq\PP{X_1>x_+}+(n-1)M_b
 \leq\delta+(n-1)M_b.
\]
Center $f_b$ without changing its variance or oscillation. On $D_n^c$,
$\widebar f_n=f_b(X_1)$, and hence stationarity gives
\begin{align*}
 \Var{\widebar f_n}
 &=\mathbb E\widebar f_n^2
 \geq\mathbb E\{f_b(X_1)^2\indc{D_n^c}\}
 \geq\Var[\pi_b]{f_b}-\osc(f_b)^2\PP{D_n},
\end{align*}
where we used $\lVert f_b\rVert_\infty\leq\osc(f_b)$.
The assumption $\osc(f_b)^2\leq R\Var[\pi_b]{f_b}$ therefore yields
\[
 \frac{\Var{\widebar f_n}}{\Var[\pi_b]{f_b}}
 \geq1-R\{\delta+(n-1)M_b\}.
\]
For $n\leq1+\lfloor\delta/M_b\rfloor$, this is at least
$1-2R\delta=(1+\varepsilon)/2>\varepsilon$. Consequently,
\[
 N_{\mathrm{PM},b}(f_b;\varepsilon)
 >1+\left\lfloor\frac{\delta}{M_b}\right\rfloor.
\]
Finally, the argument of $\widebar\Phi$ in $M_b$ satisfies
\[
 \frac{\theta^2-\sqrt a\,\theta q_{\delta/2}-a\log(2/\delta)}
      {\sqrt{\theta^2+\lambda^2}}
 =\theta-\sqrt a\,q_{\delta/2}+O(\theta^{-1})
\]
uniformly in $a\in[a_0,1)$. Mills' ratio gives
$-\log M_b=\theta^2/2+O\{\theta+\log(1+\theta)\}$, which proves the
claimed lower bound.
\end{proof}

\begin{remark}[Zero spectral gap]
Fix $b>0$ and $a\in(0,1)$, and let $r_b(x)$ denote the upper bound in
\eqref{eq:filedrawer:move-bound}. This function is increasing and tends
to zero as $x\to-\infty$. Thus, for $B_M=(-\infty,-M)$ and $x\in B_M$,
\[
 K_{{\rm PM},b}(x,B_M^c)
 \leq p_{\rm move}(x)\leq r_b(x)\leq r_b(-M).
\]
The Rayleigh quotient of $\indc{B_M}$ therefore satisfies
\[
 \frac{\calE_{K_{{\rm PM},b}}(\indc{B_M},\indc{B_M})}
      {\Var[\pi_b]{\indc{B_M}}}
 =
 \frac{\int_{B_M}\pi_b(\rd x)K_{{\rm PM},b}(x,B_M^c)}
      {\pi_b(B_M)\{1-\pi_b(B_M)\}}
 \leq\frac{r_b(-M)}{1-\pi_b(B_M)}
 \longrightarrow0
 \qquad(M\to\infty),
\]
since $\pi_b(B_M)\to0$. The variational characterization of the spectral
gap gives $\Gap(K_{{\rm PM},b})=0$.
\end{remark}

\subsection{Proof of Theorem~\ref{thm:filedrawer-cpm-cost}}
\label{sec:filedrawer-cpm-proof}

\begin{proof}[Proof of Theorem~\ref{thm:filedrawer-cpm-cost}]
Write $a=a_b$, let $\mu=p\otimes m$, and denote the augmented CPM kernel by
$K_b$. Its target is $\widetilde\pi_b=\mu(\cdot\mid A_b)$, and it is
obtained by censoring $P_b=Q_b\otimes R_{\rho_b}$ to $A_b$.
The Gaussian RWM kernel $Q_b$ is $p$-reversible and positive
semidefinite, as used in the proof of Theorem~\ref{thm:filedrawer-cost}.
The autoregressive kernel $R_{\rho_b}$ is $m$-reversible and positive
semidefinite because $R_{\rho_b}=R_{\sqrt{\rho_b}}^2$.
Hence $P_b$ and its censored kernel $K_b$ are reversible and positive
semidefinite; the latter follows from
\[
 \mu(A_b)\langle F,K_bF\rangle_{\widetilde\pi_b}
 =\langle\indc{A_b}F,P_b(\indc{A_b}F)\rangle_\mu
  +\int_{A_b}F(z)^2P_b(z,A_b^c)\,\mu(\rd z)\geq0.
\]

Use the rotation $(S,T)$ and $E_b=b(S-b)$ from the proof of
Theorem~\ref{thm:filedrawer-cost}. Under $\widetilde\pi_b$,
$T\sim\N(0,1)$ is independent of $E_b$, and
$\PP{E_b>r}\leq e^{-r}$ by~\eqref{eq:filedrawer:excess-tail}.
For fixed $\delta\in(0,1/2]$, set
\[
 L=\log(2/\delta),\qquad M=\sqrt{2\log(4/\delta)},\qquad
 H_b=\{0<E_b\leq L,\ |T|\leq M\}.
\]
Then $\widetilde\pi_b(H_b)\geq1-\delta$. Let $K_{b,H}$ be $K_b$
censored to $H_b$. In coordinates $(e,t)=(E_b,T)$,
\[
 x=\sqrt a\,(b+e/b)+\sqrt{1-a}\,t,\qquad
 u=\sqrt{1-a}\,(b+e/b)-\sqrt a\,t,
\]
with Jacobian $\rd x\,\rd u=b^{-1}\rd e\,\rd t$.
For endpoints in $H_b$ with $|t'-t|\leq b^{-1}$, both $\Delta x$ and
$\Delta u$ are $O(b^{-1})$, while $x,u=O(b)$.
Also $h_b\asymp b^{-1}$,
$\sqrt c/b\leq\sigma_b:=\sqrt{1-\rho_b^2}\leq\sqrt{2c}/b$,
and $(1-\rho_b)u=O(b^{-1})$.
Thus the base Metropolis acceptance and both standardized Gaussian
proposal densities are bounded below, uniformly over
$a\in[a_0,1)$ and $b\geq\max\{1,\sqrt c\}$. The continuous transition
density satisfies
\begin{equation*}
 k_{b,H}((e,t),(e',t'))
 =\frac{\varphi(\Delta x/h_b)\alpha(x,x')
         \varphi\{(\Delta u+(1-\rho_b)u)/\sigma_b\}}{b h_b\sigma_b}
 \geq\beta_\delta b,\qquad\beta_\delta>0,
\end{equation*}
whenever $|t'-t|\leq b^{-1}$. Both endpoints lie in $A_b$, so there
is no further selection factor. If $|t_1-t_2|\leq(2b)^{-1}$, their
destination intervals $[t_i-b^{-1},t_i+b^{-1}]\cap[-M,M]$ overlap
in length at least $(2b)^{-1}$. Integrating the common density bound
over this interval and $e'\in(0,L]$ gives overlap at least
$\alpha_\delta:=\beta_\delta L/2>0$.

We now reuse the conductance argument in the proof of
Theorem~\ref{thm:filedrawer-cost}. For any cut, define the
low-crossing sets as before, with threshold $\alpha_\delta/4$.
The overlap bound implies that their $T$-projections,
say $P_1$ and $P_2$, are separated by at least $(2b)^{-1}$.
Let $\eta_b=\widetilde\pi_b(\cdot\mid H_b)$ and let $\nu_T$
be its $T$-marginal. Since $\nu_T$ is a standard Gaussian
truncated to $[-M,M]$, it is log-concave with variance at
most one. If $A_0,B_0$ denote the two low-crossing sets,
then
\[
 \eta_b(H_b\setminus(A_0\cup B_0))
 \geq \nu_T([-M,M]\setminus(P_1\cup P_2)).
\]
Applying the separator inequality~\eqref{eq:cheeger_ineq}
to $\nu_T$ and repeating the two-case crossing-flow argument
from Theorem~\ref{thm:filedrawer-cost} gives
\begin{equation*}
 \Phi_{K_{b,H}}\geq
 \min\left\{\frac{\alpha_\delta}{8},
             \frac{\alpha_\delta}{32\sqrt3\,b}\right\},
 \qquad
 \Gap(K_{b,H})\geq\frac{\Gamma_\delta}{b^2},
 \quad\Gamma_\delta>0,
\end{equation*}
where the second bound uses Cheeger's inequality
\citep[Theorem~3.5]{lawler1988bounds}.

The inequality~\eqref{eq:var_g_bound} and restriction of the
Dirichlet form to $H_b\times H_b$ show, for bounded $F$,
\[
 \Var[\widetilde\pi_b]{F}
 \leq\frac{b^2}{\Gamma_\delta}\calE_{K_b}(F,F)
       +\delta\osc(F)^2.
\]
Since $K_b$ is positive semidefinite, applying Lemma~\ref{lem:defective-mc}
with $F(x,u)=f_b(x)$ yields
\[
 \Var{\widebar f_n}
 \leq\frac{2b^2}{n\Gamma_\delta}\Var[\pi_b]{f_b}
       +\delta\osc(f_b)^2.
\]
Taking $\delta=\min\{1/2,\varepsilon/(2R)\}$ proves
$N_{\mathrm{CPM},b}(f_b;\varepsilon)\leq\left\lceil\frac{4b^2}{\varepsilon\Gamma_\delta}\right\rceil$.
\end{proof}

\section{Additional numerical results}
\label{app:additional-numerics}
\subsection{Sampling validation}
\label{app:sampling-validation}

\begin{figure}[htbp]
\centering
\includegraphics[width=0.8\linewidth]
{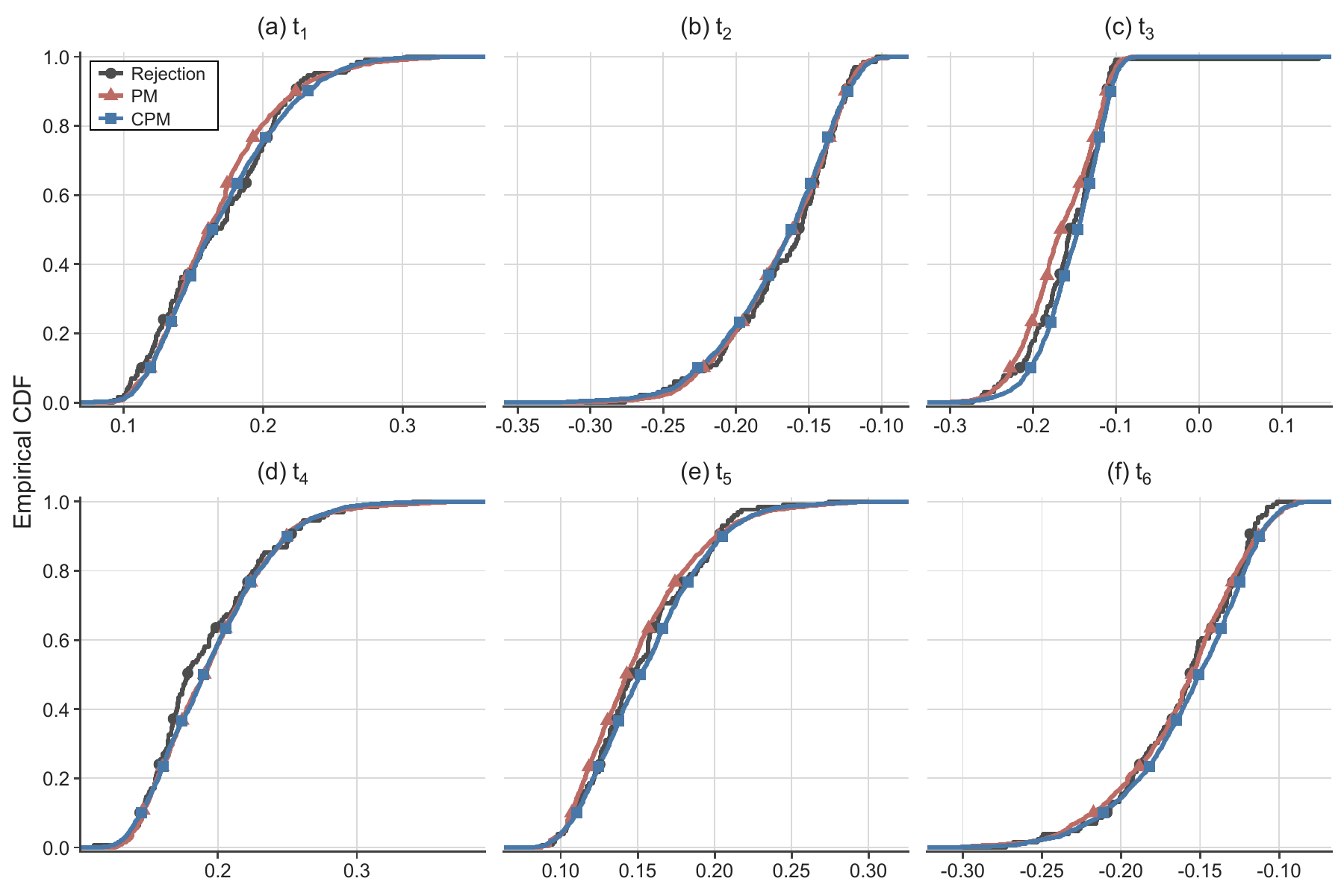}
\caption{Marginal empirical CDFs for the CV-lasso benchmark,
using $129$ rejection draws and $19{,}500$ production states
from each of PM and CPM, with repeated states retained.
Coordinates follow the variable order
$\{1,7,15,30,81,100\}$.}
\label{fig:cv-sampling-validation}
\end{figure}

\begin{figure}[htbp]
\centering
\includegraphics[width=0.8\linewidth]
{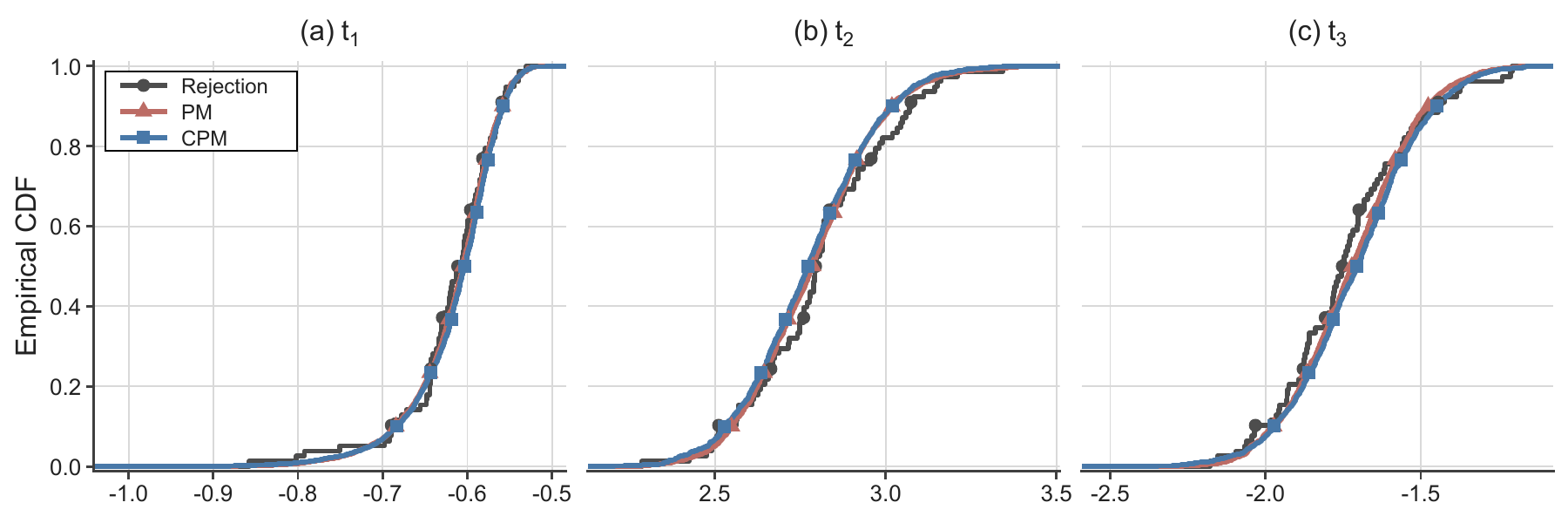}
\caption{Marginal empirical CDFs for the changepoint benchmark,
using $78$ rejection draws and $19{,}500$ production states
from each of PM and CPM, with repeated states retained.
Coordinates are successive right-minus-left segment-mean
contrasts induced by $\{71,99,140\}$.}
\label{fig:cp-sampling-validation}
\end{figure}

We provide further diagnostic checks for PM and CPM
in the sampling-efficiency studies of
Sections~\ref{sec:applications-cvlasso}
and~\ref{sec:applications-changepoint}.
Table~\ref{tab:sampling-validation-diagnostics}
summarizes the precision of coordinate-mean estimates.
Coordinate-wise ESS and MCSE are computed on the
untransformed production draws using
\texttt{posterior::ess\_basic} with chain splitting
and \texttt{posterior::mcse\_mean}, respectively.
The same ESS estimator is used in the
sampling-efficiency comparisons.
Figures~\ref{fig:cv-sampling-validation}
and~\ref{fig:cp-sampling-validation} show broadly similar
marginal empirical distribution functions.
\begin{table}[ht]
\centering
\begin{tabular}{llrr}
\toprule
Application & Method & Min. mean ESS & Max. mean MCSE \\
\midrule
CV-lasso
& PM  & 58.1  & 0.00548 \\
& CPM & 153.0 & 0.00352 \\
\midrule
Changepoint
& PM  & 239.0 & 0.01197 \\
& CPM & 309.0 & 0.01153 \\
\bottomrule
\end{tabular}
\caption{Coordinate-mean diagnostics for the sampling benchmarks.
The minimum mean ESS and maximum mean MCSE are taken
separately over all coordinates.}
\label{tab:sampling-validation-diagnostics}
\end{table}

\subsection{Finite-chain diagnostics for selective inference}
\label{app:inference-diagnostics}

We assess the Monte Carlo precision of selective
$p$-values using the test-tail indicators
\[
I_s=\mathbf1\{f(T_s)\geq f(t_{\mathrm{obs}})\},
\qquad s=1,\ldots,N_{\mathrm{MC}},
\]
where $f$ is the statistic used in the corresponding test.
Each sequence contains $N_{\mathrm{MC}}=5{,}000$
ordered production states following $200$ adaptive
warm-up iterations, with all repeated states retained.
The same ESS and MCSE estimators as above are applied
to each tail-indicator sequence.
Table~\ref{tab:inference-diagnostics} summarizes
the resulting indicator ESS and $p$-value MCSE.
All empirical $p$-values are retained in the
inference results.
At $\delta=0$, the proposed method's directional tests
reuse the corresponding quadratic-null production trajectories.

\begin{table}[ht]
\centering
\begin{tabular}{lllrr}
\toprule
Application & Experiment & Method
& Indicator ESS & $p$-value MCSE \\
\midrule
CV-lasso & Type I error & Proposed
& 94.4 & 0.0407 \\
& Type I error & Fixed-$\lambda$
& 122.5 & 0.0301 \\
& Power comparison & Proposed
& 121.4 & 0.0336 \\
& Power comparison & Extra conditioning
& 60.7 & 0.0422 \\
\midrule
Changepoint & Type I error & Proposed
& 247.8 & 0.0278 \\
& Type I error & Fixed-$K$
& 320.9 & 0.0150 \\
& Power comparison & Proposed
& 241.3 & 0.0230 \\
\bottomrule
\end{tabular}
\caption{Finite-chain diagnostics for selective $p$-values.
The Type I error experiments use quadratic test
statistics with $200$ chains per method, while the power
comparisons use directional statistics across five signal
strengths, including $\delta=0$, with $1{,}000$ chains
per method.
All reported ESS and MCSE values are medians across chains.}
\label{tab:inference-diagnostics}
\end{table}

\bibliographystyle{apalike}
\bibliography{main}

\end{document}